\documentclass[pra,twocolumn,aps,showpacs,superscriptaddress,longbibliography]{revtex4-2}

\usepackage{mypreamble}

\usepackage{overpic}
\usepackage[normalem]{ulem}
\usepackage{orcidlink}

\newcommand{\jami}{Choi-Jamiołkowski}
\newcommand{\choip}{\langle \text{Tr}[\mathcal{D}(t)^2] \rangle_{\mathrm{Haar}, T}}
\newcommand{\statep}{\overline{\mathcal{P}}}

\newcommand{\mlsr}{\ev{\tilde{r}}}

\newcommand{\unam}{Universidad Nacional Aut\'onoma de M\'exico, Ciudad de M\'exico 04510, Mexico}
\newcommand{\ifunam}{Instituto de F\'{\i}sica, \unam}
\newcommand{\icn}{Instituto de Ciencias Nucleares, \unam}

\newcommand{\pcsadd}{Center for Theoretical Physics of Complex Systems, Institute for Basic Science (IBS), Daejeon 34126, Republic of Korea}

\usepackage[draft,inline,nomargin]{fixme}
\fxsetup{theme=color}

\definecolor{jacolor}{RGB}{200,40,0}
\FXRegisterAuthor{ja}{aja}{\color{violet}JA}
\FXRegisterAuthor{mg}{amg}{\color{red}Miguel}
\FXRegisterAuthor{cd}{acd}{\color{green!50!black}Charly}

\makeatletter
\def\@lox@prtc{\section*{\@fxlistfixmename}\begingroup\def\@dotsep{4.5}}
\def\@lox@psttc{\endgroup}
\makeatother

\hypersetup{
  colorlinks=true,
  linkcolor=blue,
  filecolor=blue,      
  citecolor=blue,
  urlcolor=blue,
  pdftitle={},
  pdfauthor={}
}

\graphicspath{{figures}}

\begin{document}

\title{Dynamical irreversibility and local decoherence in quantum many-body chaos}

\author{Jose Alfredo de Leon \orcidlink{0000-0003-0045-9017}}\email{deleongarrido.jose@gmail.com}\affiliation{\ifunam}

\author{Miguel Gonzalez\,\orcidlink{0009-0004-0112-5988}}
  \email{miguel.gonzalez@correo.nucleares.unam.mx}
  \affiliation{\icn}
  \affiliation{\pcsadd}
    
\author{Carlos Diaz-Mejia \orcidlink{0000-0002-1942-8125} }\email{carlos.diaz@correo.nucleares.unam.mx} \affiliation{\icn}

\begin{abstract}
Typical dynamical quantum chaos probes are initial-state dependent (e.g., local observables or purities) and thus may fail to capture typical decoherent behavior one expects of a subsystem. 
Quantum channels fully capture the reduced dynamics of a subsystem.
Here, we investigate the purity of the Choi state of a single spin in a chain, which acts as the state representation of the channel encoding the reduced dynamics. Operationally, we show this quantity functions as an echo protocol, termed the \textit{Choi echo}. It measures the environment's recovery fidelity when subjected to a forward evolution, a completely depolarizing operation on the local subsystem, and a subsequent backward evolution.
We investigate the equilibration value of the Choi echo across the integrability-to-chaos transition in three paradigmatic spin-$1/2$ chains. We show that average single-spin decoherence does not uniquely correspond to spectral chaos. Specifically, coherent transport in integrable systems can mimic the mean relaxation of a single spin typically induced by chaotic scrambling, generating false positives for spectral chaos. This work offers a perspective bridging quantum information tools with many-body quantum chaos.
\end{abstract}
\keywords{}
\pacs{}
 
\maketitle

\section{Introduction}

The characterization of quantum chaos in many-body systems is central to understanding thermalization, ergodicity, information scrambling, and the emergence of statistical mechanics from isolated unitary dynamics~\cite{dalessio_2016_quantum}. A fundamental paradigm for diagnosing this chaotic regime stems from the Bohigas-Giannoni-Schmit conjecture~\cite{bohigas_1984_characterization}, which has been broadly generalized to assert that chaotic systems exhibit universal spectral signatures governed by Random Matrix Theory (RMT), such as level repulsion in the distribution of consecutive level spacing ratios~\cite{oganesyan_2007_localization, atas_2013_distribution}. 
While foundational, evaluating this distribution of ratios alongside other spectral probes---such as the spectral form factor, number variance, and higher-order ratios~\cite{mehta_2004_random,tekur_2020_symmetry,haake_2001_quantum}---demands, in principle, measurements of the complete many-body system. This requirement quickly becomes computationally and experimentally prohibitive as the Hilbert space dimension scales exponentially~\cite{roushan2017spectroscopic, dong_2025_measuring}.
Offering a practical pathway around this bottleneck, recent studies in spin chains have demonstrated that single-spin dynamical indicators can resolve the global spectral properties of the full system~\cite{mirkin_2021_quantum, mirkin_2021_sensing, vallejo-fabila_2025_singlesite}.

While previously studied single-spin indicators---namely purity degradation~\cite{mirkin_2021_quantum}, accumulated geometric phase~\cite{mirkin_2021_sensing}, and spin autocorrelation functions~\cite{vallejo-fabila_2025_singlesite}---capture essential facets of single-site relaxation, they inherently probe the dynamics from a restricted, trajectory-dependent point of view. To characterize the dynamics independently of the initial state, the quantum channel formalism provides the complete picture of the reduced subsystem~\cite{breuer_2002_theory,nielsen_2010_quantum,bengtsson_2006_geometry}. Indeed, under unitary evolution governed by a Hamiltonian drawn from a unitarily invariant random ensemble, the average reduced dynamics has been shown to act as depolarizing channel~\cite{znidaric_2011_nonmarkovian}---a map that isotropically contracts the Bloch sphere. Naturally, a valid question arises: does this decoherence univocally signal spectral chaos in physical systems? We address this in this work.  

To characterize a quantum channel through a single, experimentally accessible scalar, we evaluate the purity of its Choi state---the state representation of the channel in a doubled Hilbert space. This scalar distills the coherence-preserving character of the reduced dynamics~\cite{wallman_2015_estimating}. Since standard quantum process tomography carries a prohibitive experimental overhead for generic many-body systems~\cite{huang_2020_predicting}, the Choi purity is uniquely advantageous as it can be obtained via more scalable methods. For instance, it can be extracted directly by implementing a swap test on the Choi state~\cite{rosgen_2011_testing,yuan_2021_universal}---reminiscent of techniques already deployed for subsystem purity in ultracold atoms~\cite{islam_2015_measuring, kaufman_2016_quantum}---or evaluated via established randomized benchmarking protocols to estimate unitarity~\cite{wallman_2015_estimating}.

We establish a rigorous operational interpretation for the purity of the Choi state by framing it as a three-step echo protocol. Specifically, it measures the exact fidelity with which the effective environment---the entire system excluding the subsystem of interest---returns to its initial state after a sequence of forward time evolution, a local depolarizing operation that severs quantum entanglement between the subsystem and environment, and a subsequent backward time evolution. Because this metric directly gauges the stability of the environmental state against a local loss of information during this cycle, we term it the \textit{Choi echo}.

We investigate the single-spin Choi echo across the integrability-to-chaos transition in three paradigmatic chains: the mixed-field Ising model, the random-field Heisenberg chain, and the \textit{XXZ} model with a local defect. The subsystem is the leftmost spin of an open chain of length $L$, while the remaining $L-1$ spins form the environment. To ensure environmental initial-state independence, we analytically Haar average the echo over random environmental product states. Finally, its long-time equilibration value is compared against the mean level spacing ratio and the subsystem purity, which is numerically averaged over different initial states following Ref.~\cite{mirkin_2021_quantum}.

This paper is organized as follows. Section~\ref{sec:formalism} outlines the quantum channel formalism, introduces the Choi state purity, and establishes its relationship with subsystem purity. Section~\ref{sec:interpretation} develops the operational interpretation of the Choi echo. Section~\ref{sec:chaos_probe} details the spectral and dynamical metrics used as benchmarks. Section~\ref{sec:models_results} presents the spin chain models and the comparative numerical analysis, expanding upon mixed-field Ising parameter space from Ref.~\cite{mirkin_2021_quantum}; here, we show that the Choi echo provides a sharper parameter-space resolution than state purity, while demonstrating that spin excitation transport in the \textit{XXZ} with a local defect chain forces both metrics to identical values across chaotic and integrable regimes. Finally, Sec.~\ref{sec:conclusions} summarizes our findings and outlines future directions.

\section{Mathematical Framework} \label{sec:formalism}

\subsection{Quantum channels}\label{sec:reduced_dynamics}

We describe the reduced dynamics of a quantum subsystem via the quantum channel formalism. Let $\mcE_t: \mathcal{B}(\mathcal{H}_S) \to \mathcal{B}(\mathcal{H}_S)$ denote the completely positive and trace-preserving (CPTP) map governing the evolution of the subsystem density matrix $\rho_S$ at time $t$~\cite{nielsen_2010_quantum,bengtsson_2006_geometry}.

We consider a closed quantum system partitioned into a subsystem $S$ and an environment $E$. The global system evolves under a unitary operator $U$, which may arise either from continuous-time Hamiltonian dynamics or from a discrete sequence of gates. Assuming the total system is initialized in a product state $\rho_{SE}(0)=\rho_S\otimes\rho_E$, the reduced map $\mathcal{E}_t$ is obtained by tracing out the environment:
\begin{equation}\label{eq:channel_def}
\rho_S(t)=\mathcal{E}_t(\rho_S)=\operatorname{Tr}_E\!\left[\,U(\rho_S\otimes\rho_E)U^\dagger\,\right].
\end{equation}
Although the global evolution $U$ is reversible, the reduced dynamics is generally nonunitary due to system-environment entanglement. Equation~\eqref{eq:channel_def} provides a complete specification of the quantum channel $\mathcal{E}_t$ independently of the input state and constitutes the system-environment representation of $\mathcal{E}_t$.

The quantum channel $\mcE_t$ admits an equivalent representation as a quantum state via the \jami{}  isomorphism~\cite{jamiolkowski_1972_linear,choi_1975_completely}. This duality establishes a one-to-one correspondence between any CPTP map and a positive semidefinite operator acting on a doubled Hilbert space.

Let $\mathcal{H}_{S'}\simeq\mathcal{H}_S$ be an auxiliary space, and define the maximally entangled state $\ket{\Phi^+}=d_S^{-1/2}\sum_i \ket{i}_S\otimes\ket{i}_{S'}$, where $\{\ket{i}\}$ is an orthonormal basis of $\mathcal{H}_S$. The Choi state associated with $\mathcal{E}_t$ is
\begin{equation}\label{eq:choi_def}
\mathcal{D}(t)=(\mathcal{E}_t\otimes\mathcal{I}_{S'})\!\left(\dyad{\Phi^+}\right),
\end{equation}
where $\mathcal{I}_{S'}$ is the identity channel on $S'$~\cite{bengtsson_2006_geometry}. The operator $\mathcal{D}(t)$ acts on $\mathcal{H}_S\otimes\mathcal{H}_{S'}$ and uniquely encodes all properties of $\mathcal{E}_t$. The resulting matrix $\mcD(t)$ is a density matrix, i.e., a positive semidefinite matrix of unit trace, acting on the composite space $\mcH_S \otimes \mcH_{S'}$.

The Choi state $\mcD(t)$ encodes fundamental properties of the corresponding quantum channel $\mcE_t$. 
Complete positivity and trace preservation of $\mathcal{E}_t$ translate into $\mathcal{D}(t)\ge0$ and $\operatorname{Tr}_{S'}[\mathcal{D}(t)]=\mathbb{I}_S/d_S$, respectively. 
Additionally, $\mathcal{E}_t$ is said to be \textit{unital} if it preserves the maximally mixed state, i.e., $\mathcal{E}_t(\mathbb{I}_S/d_S) = \mathbb{I}_S/d_S$, which corresponds to the condition $\operatorname{Tr}_{S}[\mathcal{D}(t)] = \mathbb{I}_{S'}/d_{S}$.
The rank of $\mcD(t)$ equals the minimum environment dimension required for a \textit{minimal dilation} of $\mcE_t$, i.e., a system-environment representation of $\mcE_t$  [see Eq.~\eqref{eq:channel_def}] with the smallest environment dimension possible. Moreover, $\mcE_t$ is entanglement breaking---erasing all quantum correlations between the system and any external system---if and only if the Choi state is separable across the $S{:}S'$ bipartition.


\subsection{Purity of the Choi state}
A key scalar associated with the Choi state is its purity,

\begin{equation}\label{eq:choi:purity}
\operatorname{Tr}[\mathcal{D}(t)^2]
=
\frac{1}{d_S^2}
\sum_{k,l}\operatorname{Tr}\!\left[\mathcal{E}_t(\dyad{k}{l})\,\mathcal{E}_t(\dyad{l}{k})\right],
\end{equation}

which quantifies the coherence-preserving character of the channel. 
The right-hand side follows directly from the definition of the Choi state in Eq.~\eqref{eq:choi_def} and expresses the purity in terms of the action of the channel on the operator basis \(\{\dyad{k}{l}\}\) associated with an orthonormal basis \(\{\ket{k}\}\) of \(\mathcal{H}_S\). It satisfies
\[
\frac{1}{d_S^2}\le\operatorname{Tr}[\mathcal{D}(t)^2]\le 1.
\]
The upper bound is achieved if and only if the reduced dynamics is unitary, while the minimum corresponds to the completely depolarizing channel that maps any input state to the maximally mixed state $\mathbb{I}_S/d_S$.

The purity of the Choi state, $\text{Tr}[\mcD(t)^2]$, formally connects to several metrics in quantum information theory. The quantification of quantum channel entropies has been axiomatically established~\cite{gour_2021_entropy}. Building on the \jami{} isomorphism, the von Neumann entropy of a channel has been defined as the entropy of its corresponding Choi state $\mcD$~\cite{chu_2022_entropy}. Following this channel-state duality, the purity of the Choi state determines the linear entropy of the quantum channel, $S_L = 1 - \text{Tr}[\mcD(t)^2]$; this quantity has been proposed as a quantifier for the information loss of a quantum channel~\cite{li_2025_quantifying}. To characterize channel decoherence, metrics such as the order-2 R\'enyi map entropy~\cite{roga_2011_entropic} and the 2-norm coherence measure of a quantum process~\cite{korzekwa_2018_coherifying} have been formulated; both quantities are evaluated via the Choi state purity. Accordingly, the purity of the Choi state is a relevant quantity within the broader context of channel entropies and other quantum information metrics.

This quantity distinguishes dynamical regimes that state-based metrics cannot. For example, a full amplitude-damping channel maps all pure inputs to a pure ground state, yielding an output-state purity $\operatorname{Tr}[\rho_S(t)^2]=1$ for any pure input, identical to that of a unitary process. However, their Choi purities differ sharply: $\operatorname{Tr}[\mathcal{D}^2]=1$ for a unitary channel and $1/2$ for full amplitude damping. The Choi purity therefore probes the process coherence rather than the purity of individual trajectories.

To clarify the relation between the Choi and output-state purities, we derive the Haar-averaged output purity over pure inputs $\ket{\psi}$. Using standard identities for Haar integration~\cite{mele_2024_introduction}, we obtain
\begin{equation}\label{eq:purity_relation}
\begin{split}
\mathbb{E}\!\left\{\operatorname{Tr}\!\left[\mathcal{E}_t(\dyad{\psi})^2\right]\right\}
&=\frac{1}{d_S(d_S+1)}\bigg(
\operatorname{Tr}\!\left[\mathcal{E}_t(\mathbb{I}_S)^2\right] \\
&\hspace{1.6cm} + d_S^2\,\operatorname{Tr}\!\left[\mathcal{D}(t)^2\right]
\bigg).
\end{split}
\end{equation}
The term $\operatorname{Tr}[\mathcal{E}_t(\mathbb{I}_S)^2]$ quantifies deviations from unitality, since $\operatorname{Tr}[\mathcal{E}_t(\mathbb{I}_S)^2]\ge d_S$ with equality if and only if $\mathcal{E}_t$ is unital.

Equation~\eqref{eq:purity_relation} shows that the average state purity contains two independent contributions: the Choi purity $\operatorname{Tr}[\mathcal{D}(t)^2]$, capturing the intrinsic coherence preservation of the channel, and the unitality term, reflecting population-transfer effects. For unital channels, $\operatorname{Tr}[\mathcal{E}_t(\mathbb{I}_S)^2]=d_S$, making the averaged output purity an affine function of the Choi purity. For nonunital processes such as amplitude damping, the unitality term grows substantially, compensating for the decay of $\operatorname{Tr}[\mathcal{D}(t)^2]$ and thereby obscuring irreversibility when assessing only output-state purity. Thus, Eq.~\eqref{eq:purity_relation} clarifies why state purity alone fails to detect structural information loss in nonunital dynamics.

\begin{figure*}
\centering
\includegraphics[width=\linewidth]{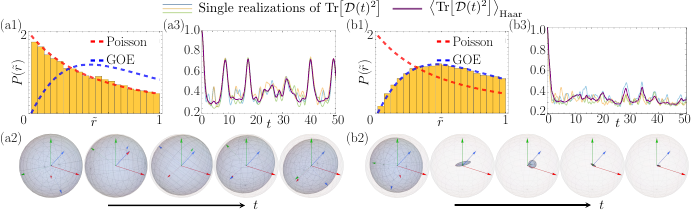}
\caption{Spectral and single-spin dynamical behavior in regular and chaotic regimes of the mixed-field Ising model [\Eref{eq:H:ising}]. Panel (a) corresponds to a regular regime [$(h_z, J) = (1.446, 0.05)$] and panel (b) to a chaotic regime [$(h_z, J) = (0.48, 0.8)$], with fixed $h_x=1$. (a1), (b1) Distribution of nearest-neighbor level spacing ratios $P(\tilde{r})$ for $L=16$ (even sector), compared to Poisson (red dashed) and GOE (blue dashed) predictions. (a2), (b2) Action of the quantum channel $\mcE_t$ on the Bloch sphere of the probe spin ($L=7$). The solid arrows represent the initial Cartesian axes of the Bloch sphere ($x, y, z$), while the dotted arrows indicate their transformed orientation at later times. In the chaotic regime, the quantum channel induces a rapid contraction of the sphere volume, contrasting with the coherent precession in the regular case. (a3), (b3) Time evolution of the Choi echo $\Tr[\mcD(t)^2]$ ($L=7$). Low-opacity lines represent single realizations with random initial product states of the environment; the thick purple line corresponds to the analytical Haar-averaged Choi echo $\mathbb{E}\{\Tr[\mathcal{D}^{2}(t)]\}$ derived in Eq.~\eqref{eq:haar:avg:choi}.}
\label{fig:cj:purity}
\end{figure*}

\section{Choi echo: operational interpretation of the Choi purity}\label{sec:interpretation}
We now establish an operational interpretation of the Choi purity [see~Eq.~\eqref{eq:choi:purity}] by showing that it plays the role of an echo quantifying recoverability under local loss of information. In this formulation, the Choi purity is not merely a static functional of the channel, but a probe of dynamical reversibility analogous in structure to an echo protocol. This perspective highlights its ability to diagnose how strongly the global evolution becomes irreversibly correlated with degrees of freedom that are inaccessible to the subsystem.

The Loschmidt echo is a fundamental tool for characterizing the stability of quantum evolution, widely employed in contexts ranging from decoherence theory to critical many-body phenomena~\cite{peres_1984_stability,gorin_2006_dynamics}. 
Defined as $L(t) = |\langle \psi_0 | e^{i (H+\Sigma) t} e^{-i H t} | \psi_0 \rangle|^2$, it quantifies the ability of a system to recover its initial state after a forward evolution under the reference Hamiltonian $H$ followed by a backward evolution governed by the perturbed Hamiltonian $H + \Sigma$.
Its decay reflects the system’s sensitivity to perturbations and serves as a canonical measure of dynamical irreversibility.

In contrast to perturbations of the Hamiltonian, the Choi purity  $\Tr[\mcD(t)^2]$ acts as an echo that probes the stability of the global evolution against a \textit{local} loss of information. Consider the channel $\mathcal{E}_t$ induced by the joint unitary $U(t)$ acting on the initially uncorrelated state $\rho_S\otimes\rho_E$, with $\rho_E=\dyad{\psi}$ any pure state. Combining the Choi definition in \Eref{eq:choi_def} with the system-environment representation of the channel in \Eref{eq:channel_def}, the purity of the Choi state can be recast in the form
\begin{equation}\label{eq:choi_as_le}
\Tr(\mcD^2) = \Bigg\langle\psi\Bigg|
\Tr_S\Bigg[
U^\dagger \Lambda_S \bqty{
U \pqty{\frac{\mathbb{I}_S}{d_S} \otimes \dyad{\psi} } U^\dagger
} U
\Bigg]
\Bigg|\psi\Bigg\rangle.
\end{equation}
Here, $\Lambda_S(\cdot) = (\mathbb{I}_S/d_S)\Tr_S(\cdot)$ represents the completely depolarizing channel acting locally on the subsystem $S$. 
For compactness, we have suppressed the explicit time arguments in this expression, denoting $U \equiv U(t)$ and $\mcD \equiv \mcD(t)$.

Equation~\eqref{eq:choi_as_le} shows that $\operatorname{Tr}[\mathcal{D}(t)^2]$ is the fidelity with which the environment returns to its initial state after a three-step protocol:  
\begin{enumerate}
\item a forward evolution under $U$, starting from a product state in which $S$ carries no information;
\item a local ``perturbation'' applied via a completely depolarizing channel $\Lambda_S$, on the system $S$ only, that breaks all entanglement between $S$ and $E$ acquired during the forward evolution; and
\item a backward evolution under $U^\dagger$.
\end{enumerate}
The Choi purity therefore quantifies the recoverability of the environment following the loss of subsystem information and plays the role of an \textit{echo}. In consequence, we refer to this quantity as the \textit{Choi echo}.

Since the Choi state provides a basis-independent representation of the channel, the Choi echo quantifies reversibility at the level of the reduced dynamics. A high echo fidelity signifies that the forward evolution has generated only weak nonlocal correlations, so that the global state remains recoverable even after the subsystem has been depolarized. Conversely, a decaying echo signals dynamical irreversibility: correlations between \(S\) and \(E\) have become sufficiently intricate that severing them at the subsystem level eliminates the possibility of reconstructing the global state. This operational viewpoint identifies the Choi echo as a tool for diagnosing irreversibility and information scrambling in many-body systems. In parallel to entropy-based indicators, which quantify locally inaccessible information, and out-of-time-ordered correlators (OTOCs), which probe operator growth, the Choi echo provides a direct measure of the recoverability of global dynamics in the presence of local information loss.

\section{Choi echo as a probe for quantum chaos}\label{sec:chaos_probe}

Diagnosing the transition to quantum chaos in many-body systems typically relies on identifying statistical fingerprints within both the energy spectrum and the system's dynamics~\cite{dalessio_2016_quantum}. Quantum-chaotic behavior manifests itself through several universal features: (i) spectral statistics governed by RMT~\cite{bohigas_1984_characterization}, (ii) eigenstate properties consistent with the eigenstate thermalization hypothesis~\cite{deutsch_1991_quantum}, and (iii) extreme sensitivity to perturbations, captured by Loschmidt echoes and OTOCs. These signatures reflect distinct facets of the same phenomenon, namely, the progressive spreading of quantum information over the many-body Hilbert space. 
In practice, however, diagnosing chaos is challenging: spectral probes require access to the bulk of the spectrum, while OTOCs and echo protocols demand controlled perturbations and precise reversals~\cite{bardarson2012unbounded,santos2010onset,alba2017entanglement,smith2016many}.
This motivates the development of diagnostics that are more computationally tractable and physically accessible.

Focusing first on spectral indicators, chaotic systems display level repulsion consistent with RMT, whereas integrable models exhibit the uncorrelated spectrum characteristic of Poisson statistics~\cite{berry_1977_level}. In this work, we use the term ``integrable'' to encompass both exactly solvable systems, e.g., via the Bethe ansatz, and systems that effectively display Poissonian spectral statistics due to finite-size effects~\protect\footnote{A rigorous definition of many-body quantum integrability requires the thermodynamic limit~\cite{caux_2011_remarks}. For instance, the mixed-field Ising model used in this work has been formally shown to be non-integrable for any non-zero $J$~\cite{chiba_2024_proof}.}. A robust measure of these short-range correlations is the mean level spacing ratio $\langle \tilde{r} \rangle$, defined as the average of $\tilde{r}_n = \min(s_n, s_{n-1})/\max(s_n, s_{n-1})$ for consecutive spacings $s_n = E_{n+1} - E_n$, evaluated within a fixed symmetry sector~\cite{oganesyan_2007_localization}. As illustrated in Figs.~\ref{fig:cj:purity}(a1) and \ref{fig:cj:purity}(b1) for the mixed-field Ising chain [see~Eq.~\eqref{eq:H:ising}], the empirical distribution shifts from the Poissonian form,
\(
P_{\text{P}}(\tilde{r}) = 2/(1+\tilde{r})^2,
\)
to the Wigner-Dyson surmised characteristic of the Gaussian orthogonal ensemble (GOE),
\(
P_{\text{GOE}}(\tilde{r}) = 8(\tilde{r} + \tilde{r}^2) / [7(1+\tilde{r}+\tilde{r}^2)^{5/2}],
\)~\cite{atas_2013_distribution}, marking the transition from integrable to chaotic behavior. Although higher-order ratios mitigate the need for explicit desymmetrization~\cite{tekur_2020_symmetry}, all spectral probes fundamentally require resolving large portions of the many-body spectrum, a task that becomes prohibitive for large Hilbert spaces~\cite{scialchi_2024_integrabilitytochaos, bernien2017probing, huang_2020_predicting, brydges2019probing}.

These limitations have motivated the search for accessible dynamical probes. Under chaotic dynamics, canonical typicality implies that small subsystems equilibrate toward stationary states that depend only on global conserved quantities~\cite{goldstein_2006_canonical, rigol_2008_thermalization}. In the context of spin chains, the degree of equilibration of a single probe spin---typically identified with the first site of the chain---has been shown to correlate strongly with spectral signatures of chaos of the entire many-body system~\cite{mirkin_2021_quantum}. Quantitatively, this is captured by the averaged subsystem purity,
\begin{equation}\label{eq:avg:P}
\overline{\mathcal{P}}=\frac{1}{N}\sum_{i=1}^N
\left(\frac{1}{T}\int_0^T \Tr[\rho_{S,i}^2(t)]\,dt\right),
\end{equation}
computed over $N$ random initial product states and a time window $[0,T]$. Across diverse spin models, $\overline{\mathcal{P}}$ exhibits a strong anticorrelation with $\langle \tilde{r} \rangle$, suggesting that single-spin decoherence can serve as an experimentally friendly probe of the integrability-to-chaos transition.

However, a fundamental question remains: does single-spin decoherence provide an unambiguous signature of many-body chaos, or can integrable dynamics produce similar behavior through mechanisms such as coherent transport? To address this, we employ the Choi echo framework introduced in Sec.~\ref{sec:interpretation}, applied to one-dimensional spin-$1/2$ chains of length $L$. We partition the system such that the first spin acts as the probe $S$, while the remaining $L-1$ spins form the environment $E$. The probe’s reduced dynamics are encoded in the quantum channel $\mathcal{E}_t$ generated by the global unitary $U(t)=e^{-iHt}$ [see~Eq.~\eqref{eq:channel_def}]. As shown in Figs.~\ref{fig:cj:purity}(a2) and \ref{fig:cj:purity}(b2), integrable evolution preserves a coherent Bloch-sphere trajectory, whereas chaotic evolution induces a rapid contraction of the Bloch volume, signaling strong decoherence. Unlike state purity, which depends on the specific initial condition, the Choi echo $\Tr[\mathcal{D}(t)^2]$ probes the intrinsic reversibility of the channel itself. By quantifying the stability of global correlations against a depolarizing operation on the probe, it provides a stricter and state-independent diagnostic of decoherence.

To isolate the generic dynamical properties of the Hamiltonian $H$ from specific initial states of the environment, we compute the analytical Haar average of the Choi echo over product states of the environment. Each environmental spin state is drawn independently from the Haar measure, a procedure corresponding to an infinite-temperature average~\footnote{When averaged over the Haar measure, a single-spin state $\dyad{\psi_j}$ becomes the maximally mixed state $\mathbb{I}/2$. The full environment state $\rho_E$, being a tensor product of $L-1$ such states, thus averages to $\langle \rho_E \rangle = \mathbb{I}_E / 2^{L-1}$, which is the state of infinite-temperature equilibrium.}. As derived in the Appendix, this yields
\begin{align}\label{eq:haar:avg:choi}
\mathbb{E}\bqty{\Tr\pqty{\mathcal{D}^{2}}} =& 
\sum_{\vec{\alpha} }
  \frac{4^{-L}}{3^{w(\vec{\alpha})}}
\Tr \Big\{ \Tr_S^2 \big[U
\qty(\sigma_1^0 \sigma_2^{\alpha_2} \ldots \sigma_L^{\alpha_L})
U^\dagger \big] \Big\},
\end{align}
where $\vec{\alpha} \in \{0,x,y,z\}^{L-1}$, and for compactness we have  suppressed the explicit time dependence. Here $\sigma_j^\mu$ denotes the Pauli 
operator labeled by $\mu \in \{0,x,y,z\}$ acting on site $j$ (with $\sigma^0 \equiv \mathbb{I}$), 
and $w(\vec{\alpha})$ counts the number of nonidentity operators in the 
environment string $\vec{\alpha} = (\alpha_2, \dots, \alpha_L)$. This analytical average is depicted in Figs.~\ref{fig:cj:purity}(a3) and \ref{fig:cj:purity}(b3), where the Haar-averaged Choi echo (thick line) effectively filters out the fluctuations inherent to single environment realizations (low-opacity lines). To construct a direct analog of $\langle \tilde{r} \rangle$ and $\overline{\mathcal{P}}$, we also average in time the Haar-averaged Choi echo:

\begin{equation}\label{eq:avg:CJ}
\expval{\Tr[\mcD(t)^2]}_{\text{Haar}, T} = \frac{1}{T}
\int_{0}^{T} \mathbb{E}\Big\{\Tr[\mathcal{D}^{2}(t)]\Big\} \, dt.
\end{equation}

The next section compares $\langle \tilde{r} \rangle$, $\overline{\mathcal{P}}$, and the Haar-averaged Choi echo across different models and parameter regimes.

\section{Model Systems and Numerical Analysis}\label{sec:models_results}

We now assess the performance of the Choi echo as a dynamical probe for the transition to quantum chaos. To this end, we analyze three paradigmatic one-dimensional spin-$1/2$ models that exemplify distinct mechanisms of integrability breaking—competing fields, quenched disorder, and local defects. This diversity provides a stringent test of whether the echo can meaningfully separate chaotic from nonchaotic dynamics using information only from single-spin dynamics.

\subsection{Spin chain Hamiltonians}\label{sec:spin_chains}

\begin{figure*}
\includegraphics[width=\textwidth]{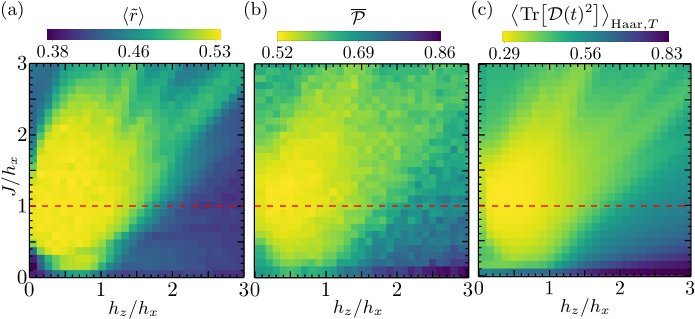}
\caption{Comparison of short-range spectral statistics and single-spin dynamical probes for the mixed-field Ising model [\Eref{eq:H:ising}] as a function of $h_z/h_x$ and $J/h_x$. (a) Mean level spacing ratio $\langle \tilde{r} \rangle$, computed for $L = 16$ within the even symmetric sector. (b), \textbf{(c)} Single-spin dynamical probes computed for the spin at the first site ($L = 7$): (b) averaged subsystem state purity $\statep$ [\Eref{eq:avg:P}] ($N=50$, $T=100$) and (c) averaged Choi echo $\choip$ [\Eref{eq:avg:CJ}] ($T=100$). The horizontal red dashed line ($J/h_x=1$) indicates the parameter cut where the correspondence between $\mlsr$ and $\statep$ was analyzed in Ref.~\cite{mirkin_2021_quantum} for $h_z/h_x \in [0, 2.5]$.}
\label{fig:cj:purity:ising}
\end{figure*}

\subsubsection{Mixed-field Ising model}\label{sec:ising}
We first consider the mixed-field Ising model with open boundary conditions, described by the Hamiltonian:
\begin{equation}\label{eq:H:ising}
H = \sum_{i=1}^L \left( h_x \sigma_i^x + h_z \sigma_i^z \right) - J \sum_{i=1}^{L-1} \sigma_i^z \sigma_{i+1}^z \, ,
\end{equation}
where $\sigma_i^\alpha$ ($\alpha \in \{x,y,z\}$) are Pauli operators, $L$ is the chain length, $J$ is the nearest-neighbor Ising interaction, and $h_x, h_z$ are transverse and longitudinal field strengths. This model possesses a global spatial reflection symmetry $\mathcal{R}$ ($i \leftrightarrow L-i+1$), decomposing the Hilbert space into even- and odd- parity sectors. The model is integrable in two limits: the transverse-field Ising model ($h_z = 0$), solvable via Jordan-Wigner transformation~\cite{sachdev_2011_quantum}, and the so-called classical Ising model ($h_x = 0$)~\cite{baxter_2007_exactly}. When $J=0$, the system decouples into a noninteracting set of $L$ spins. The simultaneous presence of all three terms breaks integrability, rendering it a standard testbed for quantum chaos~\cite{kim_2014_testing,garrison_2018_does,borgonovi_2016_quantum}. Notably, it has been rigorously proven that for any nonzero $J$, $h_x$, and $h_z$, the model possesses no nontrivial local conserved quantities~\cite{chiba_2024_proof}. The chaotic regime is frequently studied using parameters where all three terms are of comparable magnitude, $J \sim h_x \sim h_z$, as this maximizes the noncommutation responsible for driving chaotic behavior~\cite{karthik_2007_entanglement,atas_2017_quantum,camargo_2024_spread}.

\subsubsection{Heisenberg model with random fields}\label{sec:heisenberg}
Next, we examine the Heisenberg model with random fields:
\begin{equation}\label{eq:H:heisenberg}
H = \frac{1}{4}\sum_{i=1}^{L-1} \left(\vec{\sigma}_i \cdot \vec{\sigma}_{i+1}\right) + \frac{1}{2}\sum_{i=1}^L h_i \sigma_i^z \, ,
\end{equation}
where $h_i$ are independent random variables drawn uniformly from the interval $[-h, h]$, and $h$ denotes the disorder strength. This model conserves the total magnetization $M_z = \sum_i \sigma_i^z$. This $U(1)$ symmetry is equivalent to conserving the total number of up spins, $N_{\uparrow}$, related to the magnetization via $\ev{M_z} = 2N_{\uparrow} - L$. The clean limit ($h=0$) is integrable by Bethe-anzatz~\cite{bethe_1931_zur}. For $h>0$, integrability is broken. This Hamiltonian describes the transition from an ergodic or thermal phase ($h \lesssim 3.5$) to a many-body localized (MBL) phase at strong disorder~\cite{oganesyan_2007_localization,luitz_2015_manybody,abanin_2019_colloquium}.

\subsubsection{\textit{XXZ} model with a local defect}\label{sec:xxz}
Finally, we study the \textit{XXZ} model with a single local defect:
\begin{equation}\label{eq:H:xxz}
H = \frac{1}{4}\sum_{i=1}^{L-1} \left[ J_{xy} (\sigma_i^x\sigma_{i+1}^x + \sigma_i^y\sigma_{i+1}^y) + J_z \sigma_i^z\sigma_{i+1}^z \right] + \frac{1}{2}\varepsilon \sigma_d^z \, .
\end{equation}
Here, $J_{xy}$ and $J_z$ are anisotropic interaction strengths, and $\varepsilon$ is the field strength at a defect site $d$. Like the Heisenberg model, this system conserves $M_z$. The defect is placed near the center to break spatial reflection symmetry explicitly. For $\varepsilon = 0$, the model is integrable~\cite{bethe_1931_zur}. A nonzero defect $\varepsilon \neq 0$ breaks integrability~\cite{santos_2004_integrability,brenes_2018_hightemperature,pandey_2020_adiabatic}, providing a minimal model where chaos is induced solely by a local perturbation on a single site. We note that in the limit $\varepsilon \gg 1$, the chain effectively decouples into two noninteracting parts.

\subsection{Results}\label{sec:results_discussion}
We benchmark the Haar- and time-averaged Choi echo $\choip$ [Eq.~\eqref{eq:avg:CJ}] against two reference indicators:  
(i) the averaged subsystem state purity $\statep$ [Eq.~\eqref{eq:avg:P}] and  
(ii) the mean level spacing ratio $\langle \tilde{r} \rangle$, which provides the spectral ground for quantum many-body chaos. While both purities are computed for small chains ($L=7$) with the probe at site $1$, the spectral statistics are evaluated via exact diagonalization on significantly larger systems ($L=16$ or $18$). This mismatch in system sizes imposes a stringent requirement: a reliable single-spin dynamical probe must be robust to finite-size effects and must recover the chaotic transition inferred from the spectral benchmark.

\subsubsection{Resolving chaos and decoupling in the mixed-field Ising model}
\Fref{fig:cj:purity:ising} compares the three quantities across the $(J/h_x,\,h_z/h_x)$ plane. The mean level spacing ratio is computed for $L=16$ within the even sector (dimension 32,898). As shown in Fig.~\ref{fig:cj:purity:ising}(a), it exhibits a well-defined chaotic dome where $\langle \tilde{r} \rangle \approx 0.53$. Despite the reduced system size, both $\statep$ and $\choip$ reproduce this structure, with low purity correlating with high $\langle \tilde{r} \rangle$. Notably, the Choi echo [Fig.~\ref{fig:cj:purity:ising}(c)] resolves the boundaries of the chaotic region more sharply than the state purity [Fig.~\ref{fig:cj:purity:ising}(b)].

However, a discrepancy emerges in the regular region $1.5 \le h_z/h_x \le 3$ under weak coupling $J/h_x < 0.5$, indicating that the dynamical probes do not anticorrelate with the spectral statistics across the full parameter space. In this regime, the mean level spacing ratio $\mlsr$ saturates at its Poisson value ($\approx 0.38$), whereas both dynamical probes display a smooth gradient. As $J/h_x$ decreases, the Choi echo and the state purity increase gradually, reaching their maxima only when $J \to 0$. Thus, while spectral statistics detect a uniform regular phase, single-spin probes remain sensitive to the competition between the Ising coupling $J$ and the local fields $(h_x, h_z)$. The Choi echo, in particular, captures the continuous drift toward locally unitary dynamics as interactions become negligible.

Finally, in regions where the interaction $J$ is comparable to the local fields, single-spin probes regain a faithful anticorrelation with $\mlsr$, as observed along the parameter cut indicated by the red dashed line in Fig.~\ref{fig:cj:purity:ising} ($J/h_x = 1$). This is precisely the regime analyzed in Ref.~\cite{mirkin_2021_quantum} for $0 < h_z/h_x < 2.5$, where both dynamical and spectral indicators show consistent signatures of chaos.

\subsubsection{Tracking the thermal-to-MBL transition in the Heisenberg model with random fields}

\begin{figure}
\includegraphics[width=\columnwidth]{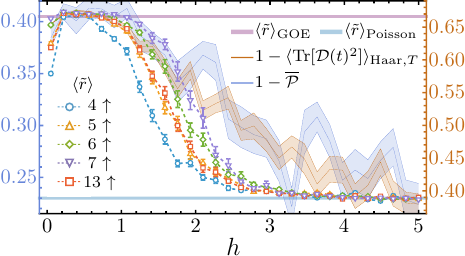}
\caption{Comparison of spectral statistics and single-spin dynamical probes for the Heisenberg model with random fields [\Eref{eq:H:heisenberg}] as a function of disorder strength $h$. All quantities are averaged over 30 disorder realizations. Symbols: Mean level spacing ratio $\langle \tilde{r} \rangle$, computed for $L=18$ within different magnetization sectors $N_{\uparrow}$ (horizontal lines indicate GOE and Poisson limits). Solid lines: Single-spin dynamical probes computed for the spin at the first site ($L=7$), plotted as deviations from unity (impurities). Left axis (blue): averaged subsystem state impurity $1-\statep$ [\Eref{eq:avg:P}] ($N=50$, $T=100$); Right axis (brown): averaged Choi echo deviation $1-\choip$ [\Eref{eq:avg:CJ}] ($T=100$). Shaded bands and error bars indicate the standard deviation.}
\label{fig:cj:purity:heisenberg}
\end{figure}

In the Heisenberg model with random fields, we examine the transition from the thermal to the MBL phase. Figure~\ref{fig:cj:purity:heisenberg} displays the spectral and dynamical metrics as a function of disorder strength $h$, with all quantities averaged over 30 disorder realizations. The mean level spacing ratio $\mlsr$ is computed for $L=18$ across five magnetization sectors, specified by the number of up spins $N_{\uparrow} \in \{4, 5, 6, 7, 13\}$, with corresponding Hilbert-space dimensions 3060, 8568, 18564, 31824, and 8568, respectively. These results, shown as symbols with error bars, capture the crossover from GOE to Poisson statistics around $h_c \approx 3.5$. Here, we observe a robust agreement between the spectral statistics and the single-spin probes. Both the deviation of the Choi echo from unity, $1-\choip$, and the state impurity ($1-\statep$) remain high in the thermal phase ($h \lesssim 1$) and decay as the system transitions to the MBL phase. Significantly, in the deep thermal phase, the Choi echo exhibits negligible variance across disorder realizations (the standard deviation band in Fig.~\ref{fig:cj:purity:heisenberg} is virtually indiscernible), suggesting it is a stable quantifier of the dynamical map's scrambling power. In contrast, the state purity shows larger fluctuations, reflecting its dependence on the specific initial-state trajectory.

\subsubsection{Transport-induced false positives for chaos in the \textit{XXZ} model with a local defect}\label{sec:results_xxz}

\begin{figure*}
\includegraphics[width=\textwidth]{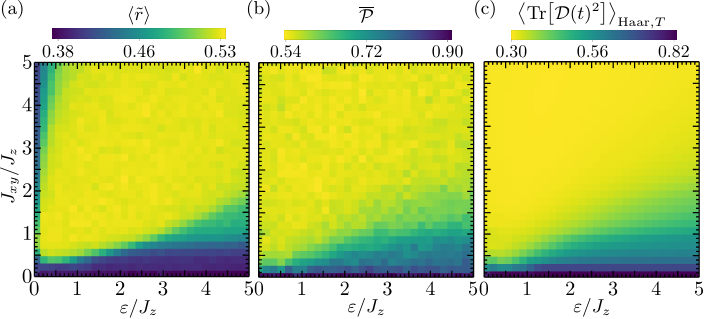}
\caption{Comparison of spectral statistics and single-spin dynamical probes for the \textit{XXZ} model with a local defect [\Eref{eq:H:xxz}] as a function of $J_{xy}/J_z$ and $\varepsilon/J_z$. (a) Mean level spacing ratio $\langle \tilde{r} \rangle$, computed for $L=18$ ($N_{\uparrow}=7$) with the defect at site $d=9$. (b), \textbf{(c)} Single-spin dynamical probes computed for the spin at the first site ($L=7$) with the defect at site $d=3$: (b) averaged subsystem state purity $\statep$ [\Eref{eq:avg:P}] ($N=50$, $T=100$) and (c) averaged Choi echo $\choip$ [\Eref{eq:avg:CJ}] ($T=100$). In both configurations, the defect placement explicitly breaks spatial reflection symmetry.}
\label{fig:cj:purity:xxz}
\end{figure*}

Finally, the results for the \textit{XXZ} model with a local defect reveal a second fundamental limitation of the single-spin dynamical probes when contrasted with spectral statistics of the full chain. Figure~\ref{fig:cj:purity:xxz} shows intensity maps of $\mlsr$, $\statep$, and $\choip$ as functions of the anisotropy $J_{xy}/J_z$ and defect strength $\varepsilon/J_z$. The spectral benchmark $\mlsr$ [Fig.~\ref{fig:cj:purity:xxz}(a)] is computed for $L=18$ in the $N_{\uparrow}=7$ sector (dimension 31824), with the defect placed at site $d=9$. The dynamical probes [Figs.~\ref{fig:cj:purity:xxz}(b) and \ref{fig:cj:purity:xxz}(c)] are obtained for $L=7$ with the defect at site $d=3$. In both cases, the defect lies in the bulk but away from the reflection axis, ensuring explicit parity breaking. This placement avoids accidental residual symmetries and makes the intensity maps qualitatively comparable despite the different system sizes.

The most prominent discrepancy appears in the region $J_{xy}/J_z > 1$ with a vanishing defect $\varepsilon \to 0$. Here the model approaches the clean \textit{XXZ} chain and is integrable, a fact captured by $\mlsr$, which saturates at the Poisson value. In sharp contrast, both dynamical probes yield low values that are indistinguishable from the chaotic regime. This constitutes a spurious signature of GOE statistics: strong interactions lead to fast single-spin relaxation of the probe, overwhelming the global integrable structure and producing a false positive for chaos.

A second, more subtle discrepancy mirrors the behavior observed in the mixed-field Ising model. For $J_{xy}/J_z \lesssim 1$, the spectral indicator $\mlsr$ exhibits a sharp transition toward regular behavior as $J_{xy}$ decreases. The dynamical probes, however, respond through a smooth gradient: as $J_{xy} \to 0$, the flip-flop term responsible for spin transport vanishes and the probe dynamics crosses over continuously to an Ising-type dephasing channel. Thus, while spectral statistics detect the abrupt breakdown of GOE correlations, the single-spin probes remain sensitive to the gradual suppression of transport rather than the transition to the integrable limit.

\subsection{Discussion}\label{sec:discussion}
The comparison across the three models yields a coherent picture of the diagnostic capabilities and limitations of single-spin dynamical probes. The Choi echo quantifies the recoverability of global correlations after information on the probe is locally erased. Its magnitude reflects how effectively the environment can rebuild the correlations required for reversibility, rather than the spectral complexity encoded in the correlations of the spectrum.

In the Heisenberg model with random fields this recoverability is intrinsically suppressed. Chaotic evolution rapidly distributes information across the system and environment, and the depolarization on the probe removes the correlations needed to reconstruct the global state. As a result, both the Choi echo and the state purity track the integrability-to-chaos crossover with high fidelity, aligning with the behavior of short-range correlations of the spectrum.

However, the \textit{XXZ} chain with a local defect exposes a key limitation: strong single-spin relaxation does not uniquely signal spectral chaos. In this model, ballistic or diffusive transport generates rapid entanglement between the probe and the rest of the chain even in parameter regimes where the many-body spectrum remains integrable. From the standpoint of the Choi echo, these transport-driven processes are operationally indistinguishable from genuine scrambling—both mechanisms disperse correlations across the system. Consequently, single-spin probes fundamentally diagnose the strength of dynamical coupling and the efficiency of information propagation, rather than the presence of correlations in the spectrum.

The mixed-field Ising model highlights a complementary advantage of the Choi echo. Because it characterizes the unitarity of the reduced dynamics instead of the mixedness of particular trajectories, it resolves the approach to the decoupled limit more sharply than state-based metrics. This distinction underscores the conceptual separation between probes of dynamical reversibility and probes of state purity: while both reflect aspects of equilibration, only the former directly quantify the intrinsic stability of the reduced dynamics under local perturbations.

From a technical standpoint, the Choi echo offers a more intrinsic diagnostic of reduced dynamics than the subsystem state purity. The analytical relationship established in Eq.~\eqref{eq:purity_relation} clarifies why state purity can remain high in nonunitary regimes, whereas the echo directly quantifies deviations from unitarity. This distinction proved essential in the mixed-field Ising model, where the echo resolved the approach to the decoupled limit with significantly higher fidelity.

Crucially, however, our results challenge the universality of the ``single-spin decoherence--spectral chaos in the entire system correspondence'' between subsystem dynamics and spectral statistics~\cite{mirkin_2021_quantum}. While chaotic dynamics (as in the random-field Heisenberg model) reliably induce high decoherence, we demonstrate that the converse does not hold. The emergence of ``false positives'' in the \textit{XXZ} model with a local defect, specifically within a parameter regime exhibiting Poissonian statistics, shows that single-spin probes cannot distinguish between genuine scrambling induced by many-body chaos and entanglement generation driven by coherent transport. In both regimes, the mechanism of information loss---from the perspective of the single-spin subsystem---is operationally identical: the environment acts as an effective sink for correlations, rendering the single-spin dynamics irreversible despite the underlying integrability of the model. We anticipate that such false positives are not unique to the \textit{XXZ} chain but will arise generically in integrable systems perturbed by local defects~\cite{santos_2020_speck}, where the conserved quantities of the integrable limit are nonlocal and are all broken by the presence of a single-site defect, as well as in integrable systems where the entangling power can mimic the effect of quantum many-body chaos~\cite{betheanzatsmagic2025}.

The combined analysis across the mixed-field Ising, random-field Heisenberg, and defected \textit{XXZ} models demonstrates both the strengths and the boundaries of single-spin dynamical indicators. While the echo reliably tracks the chaotic transition in regimes where spectral complexity and dynamical irreversibility coincide, it also reveals scenarios in which rapid single-spin relaxation arises from mechanisms unrelated to many-body chaos. In particular, coherent transport in integrable models can mimic the relaxation patterns typically associated with scrambling, underscoring that single-spin probes fundamentally quantify the efficacy of information propagation rather than the fine structure of the spectrum.

\section{Conclusions}\label{sec:conclusions}
Within the framework of reduced dynamics described by quantum channels, we establish a physically relevant operational interpretation for the purity of the Choi state. Specifically, it quantifies the recoverability of the environment---the total system minus the subsystem the channel acts upon---after a local operation destroys all system-environment entanglement generated during evolution. This perspective elevates the Choi purity from a mathematical functional to a direct probe of the interaction strength between different parts of a quantum system. In the noninteracting limit, the Choi echo reaches its maximum; as the interaction grows stronger, the echo decreases, since the environment becomes incapable of returning to its initial state after severing the quantum correlations shared with the subsystem.

When utilizing the single-spin Choi echo as a diagnostic for quantum chaos, our results show that the equilibration value of single-spin dynamics does not uniquely correspond to spectral chaoticity. In the \textit{XXZ} chain with a local defect, coherent spin transport drives a relaxation process that renders the chaotic and integrable limits operationally indistinguishable. In contrast, for the mixed-field Ising and Heisenberg with random fields chains, the Choi echo maps the chaotic boundaries in parameter space with significantly higher resolution than subsystem purity. This sharper resolution underscores the inherent advantage of characterizing the full quantum channel rather than specific state trajectories.

Our findings demonstrate that single-spin decoherence, with the environment effectively initialized in an infinite-temperature state through an analytical Haar average over product states, cannot be blindly equated with quantum chaos, thereby opening several avenues for future work. From a many-body perspective, it is critical to investigate this decoupling between single-spin relaxation and chaos in broader classes of models, particularly where the conserved quantities of the integrable regime are nonlocal. Furthermore, establishing the rigorous statistical baseline of the Choi echo requires studying its mean value and fluctuations when the evolution is sampled from a unitarily invariant ensemble. In this context, scaling the subsystem size beyond a single spin becomes particularly compelling; for larger subsystems, the average Choi purity will progressively deviate from the minimal depolarizing limit ($1/d_S^2$), potentially revealing the exact dimensional thresholds required to univocally isolate genuine many-body chaos from transport-induced artifacts.

\section*{Acknowledgments}
We acknowledge discussions with V. H. T. Brauer at early stages of this project, and valuable feedback from D. Wisniacki, C. Pineda, J. G. Hirsch, F. de Melo, and B. Dietz.
All authors acknowledge SECIHTI through funding for graduate studies.
J.A.d.L. acknowledges support by UNAM-PAPIIT Grant No. IG101324 and SECIHTI Grant No. CBF-2025-I-1548. M.G. acknowledges financial support from the Institute for Basic Science (IBS) in the Republic of Korea through Project No. IBS-R024-D1. We acknowledge the support of the Computation Center--ICN, in particular to Enrique Palacios, Luciano Díaz, and Eduardo Murrieta.



\appendix
\section{Haar average of the Choi state purity $\Tr[\mcD(t)^2]$}
\label{app:haar}
In this Appendix, we derive the Haar-expectation value of the
Choi state purity, \Eref{eq:haar:avg:choi}. The channel $\mcE_t$
acts upon the first spin ($S$) in the chain,
while the environment $E$ consists of the $L-1$ spins
$i=2, \dots, L$, prepared in a random product state
$\ket{\psi_E} = \bigotimes_{i=2}^L \ket{\phi_i}$.

The Choi echo for this channel is related to the evolved
environment state $\rho_E(t) = \Tr_S[U(t) (\mathbb{I}_S/2 \otimes
\dyad{\psi_E}) U^\dagger(t)]$ via
$\Tr[\mcD(t)^2] = \Tr_E[\rho_E(t)^2]$.
We compute the Haar average of this quantity over $\ket{\psi_E}$
using the swap trick $\Tr[A^2] = \Tr[\mathbb{S} (A \otimes A)]$:
\begin{align}
\mathbb{E}\Big\{\Tr[\mcD(t)^2]\Big\}
&= \Tr_{E_A E_B} \Big\{
\mathbb{S}_E \, \mathbb{E}\big[ \rho_E(t) \otimes \rho_E(t) \big] \Big\},
\label{eq:app:avg_swap}
\end{align}
and then:
\begin{widetext}
\begin{align}
\mathbb{E}\big[ \rho_E(t) \otimes \rho_E(t) \big]
&= \mathbb{E}\Big\{
\Tr_{S_A S_B} \Big(
U^{\otimes 2}
\qty( \frac{\mathbb{I}_S}{2} \otimes \dyad{\psi_E} )^{\otimes 2}
U^{\dagger \otimes 2}
\Big) \Big\}
%
= \Tr_{S_A S_B} \Big[
U^{\otimes 2}
\qty( \frac{\mathbb{I}_S^{\otimes 2}}{4} \otimes
\mathbb{E}\big[\dyad{\psi_E}^{\otimes 2}\big] )
U^{\dagger \otimes 2}
\Big].
\label{eq:app:avg_rhoE_tensor}
\end{align}
\end{widetext}
Due to the product state structure, the environment average
factorizes:
\begin{equation}
\mathbb{E}\big[\dyad{\psi_E}^{\otimes 2}\big]
= \bigotimes_{i=2}^L
\mathbb{E}\big[\dyad{\phi_i}^{\otimes 2}\big]
= \bigotimes_{i=2}^L \qty( \frac{\mathbb{I}_i + \mathbb{S}_i}{6} ).
\label{eq:app:haar_moment_2_qubit}
\end{equation}
Here, $\mathbb{E}[\dyad{\phi}^{\otimes 2}] = (\mathbb{I} + \mathbb{S})/(d(d+1))$
is the standard second moment \cite{collins_2022_weingarten},
which for qubits ($d=2$) yields
$(\mathbb{I} + \mathbb{S})/6$.

Substituting Eqs.~\eqref{eq:app:avg_rhoE_tensor} and \eqref{eq:app:haar_moment_2_qubit} into \Eref{eq:app:avg_swap},
\begin{align}
\mathbb{E}\Big\{\Tr[\mcD(t)^2]\Big\}
= \frac{1}{4} \Tr \Big[ & (\mathbb{I}_S^{\otimes 2} \otimes \mathbb{S}_E)
U^{\otimes 2} \nonumber \\
& \times \qty( \mathbb{I}_S^{\otimes 2} \otimes
\bigotimes_{i=2}^L \frac{\mathbb{I}_i + \mathbb{S}_i}{6} )
U^{\dagger \otimes 2}
\Big].
\end{align}
We expand the averaged projector in the Pauli basis using
$\frac{\mathbb{I} + \mathbb{S}}{6} =
\frac{1}{12} \sum_{k \in \{0,x,y,z\}} c_k (\sigma_k \otimes \sigma_k)$,
where $c_0 = 3$ and $c_{x,y,z} = 1$. The tensor product becomes
\begin{align}
\bigotimes_{i=2}^L \frac{\mathbb{I}_i + \mathbb{S}_i}{6}
&= \frac{1}{12^{L-1}} \sum_{\vec{\alpha}}
\qty(\prod_{j=2}^{L} c_{\alpha_j})
(\sigma_{\vec{\alpha}}^E \otimes \sigma_{\vec{\alpha}}^E)
\nonumber \\
&= \frac{1}{12^{L-1}} \sum_{\vec{\alpha}}
3^{L-1-w(\vec{\alpha})}
(\sigma_{\vec{\alpha}}^E \otimes \sigma_{\vec{\alpha}}^E),
\end{align}
where $\vec{\alpha} = (\alpha_2, \dots, \alpha_L) \in \{0,x,y,z\}^{L-1}$,
$\sigma_{\vec{\alpha}}^E = \bigotimes_{i=2}^L \sigma_{\alpha_i}$,
and $w(\vec{\alpha})$ counts the nonidentity operators in $\vec{\alpha}$.

Defining the full Pauli operator
$\sigma_{\vec{\alpha}}^F = \mathbb{I}_1 \otimes \sigma_{\vec{\alpha}}^E$
(using $\sigma_0 \equiv \mathbb{I}$), and inserting the
expansion, we find
\begin{widetext}
\begin{align}
\mathbb{E}\Big\{\Tr[\mcD(t)^2]\Big\}
&= \frac{1}{4} \cdot \frac{1}{12^{L-1}} \sum_{\vec{\alpha}} 3^{L-1-w(\vec{\alpha})}
\Tr \Big\{ (\mathbb{I}_S^{\otimes 2} \otimes \mathbb{S}_E)
\Big[ U (\sigma_{\vec{\alpha}}^F) U^\dagger
\otimes U (\sigma_{\vec{\alpha}}^F) U^\dagger \Big] \Big\}
\nonumber \\
&= \frac{1}{4^{L}} \sum_{\vec{\alpha} \in \{0,x,y,z\}^{L-1}}
\frac{1}{3^{w(\vec{\alpha})}}
\Tr_E \Big( \qty{ \Tr_S \big[U(t)
\sigma_{\vec{\alpha}}^F U^\dagger(t) \big] }^2 \Big).
\label{eq:app:final_result}
\end{align}
\end{widetext}


\begin{thebibliography}{64}%
\makeatletter
\providecommand \@ifxundefined [1]{%
 \@ifx{#1\undefined}
}%
\providecommand \@ifnum [1]{%
 \ifnum #1\expandafter \@firstoftwo
 \else \expandafter \@secondoftwo
 \fi
}%
\providecommand \@ifx [1]{%
 \ifx #1\expandafter \@firstoftwo
 \else \expandafter \@secondoftwo
 \fi
}%
\providecommand \natexlab [1]{#1}%
\providecommand \enquote  [1]{``#1''}%
\providecommand \bibnamefont  [1]{#1}%
\providecommand \bibfnamefont [1]{#1}%
\providecommand \citenamefont [1]{#1}%
\providecommand \href@noop [0]{\@secondoftwo}%
\providecommand \href [0]{\begingroup \@sanitize@url \@href}%
\providecommand \@href[1]{\@@startlink{#1}\@@href}%
\providecommand \@@href[1]{\endgroup#1\@@endlink}%
\providecommand \@sanitize@url [0]{\catcode `\\12\catcode `\$12\catcode
  `\&12\catcode `\#12\catcode `\^12\catcode `\_12\catcode `\%12\relax}%
\providecommand \@@startlink[1]{}%
\providecommand \@@endlink[0]{}%
\providecommand \url  [0]{\begingroup\@sanitize@url \@url }%
\providecommand \@url [1]{\endgroup\@href {#1}{\urlprefix }}%
\providecommand \urlprefix  [0]{URL }%
\providecommand \Eprint [0]{\href }%
\providecommand \doibase [0]{https://doi.org/}%
\providecommand \selectlanguage [0]{\@gobble}%
\providecommand \bibinfo  [0]{\@secondoftwo}%
\providecommand \bibfield  [0]{\@secondoftwo}%
\providecommand \translation [1]{[#1]}%
\providecommand \BibitemOpen [0]{}%
\providecommand \bibitemStop [0]{}%
\providecommand \bibitemNoStop [0]{.\EOS\space}%
\providecommand \EOS [0]{\spacefactor3000\relax}%
\providecommand \BibitemShut  [1]{\csname bibitem#1\endcsname}%
\let\auto@bib@innerbib\@empty
\bibitem [{\citenamefont {D'Alessio}\ \emph {et~al.}(2016)\citenamefont
  {D'Alessio}, \citenamefont {Kafri}, \citenamefont {Polkovnikov},\ and\
  \citenamefont {Rigol}}]{dalessio_2016_quantum}%
  \BibitemOpen
  \bibfield  {author} {\bibinfo {author} {\bibfnamefont {L.}~\bibnamefont
  {D'Alessio}}, \bibinfo {author} {\bibfnamefont {Y.}~\bibnamefont {Kafri}},
  \bibinfo {author} {\bibfnamefont {A.}~\bibnamefont {Polkovnikov}},\ and\
  \bibinfo {author} {\bibfnamefont {M.}~\bibnamefont {Rigol}},\ }\href
  {https://doi.org/10.1080/00018732.2016.1198134} {\bibfield  {journal}
  {\bibinfo  {journal} {Adv. Phys.}\ }\textbf {\bibinfo {volume} {65}},\
  \bibinfo {pages} {239} (\bibinfo {year} {2016})}\BibitemShut {NoStop}%
\bibitem [{\citenamefont {Bohigas}\ \emph {et~al.}(1984)\citenamefont
  {Bohigas}, \citenamefont {Giannoni},\ and\ \citenamefont
  {Schmit}}]{bohigas_1984_characterization}%
  \BibitemOpen
  \bibfield  {author} {\bibinfo {author} {\bibfnamefont {O.}~\bibnamefont
  {Bohigas}}, \bibinfo {author} {\bibfnamefont {M.~J.}\ \bibnamefont
  {Giannoni}},\ and\ \bibinfo {author} {\bibfnamefont {C.}~\bibnamefont
  {Schmit}},\ }\href {https://doi.org/10.1103/PhysRevLett.52.1} {\bibfield
  {journal} {\bibinfo  {journal} {Phys. Rev. Lett.}\ }\textbf {\bibinfo
  {volume} {52}},\ \bibinfo {pages} {1} (\bibinfo {year} {1984})}\BibitemShut
  {NoStop}%
\bibitem [{\citenamefont {Oganesyan}\ and\ \citenamefont
  {Huse}(2007)}]{oganesyan_2007_localization}%
  \BibitemOpen
  \bibfield  {author} {\bibinfo {author} {\bibfnamefont {V.}~\bibnamefont
  {Oganesyan}}\ and\ \bibinfo {author} {\bibfnamefont {D.~A.}\ \bibnamefont
  {Huse}},\ }\href {https://doi.org/10.1103/PhysRevB.75.155111} {\bibfield
  {journal} {\bibinfo  {journal} {Phys. Rev. B}\ }\textbf {\bibinfo {volume}
  {75}},\ \bibinfo {pages} {155111} (\bibinfo {year} {2007})}\BibitemShut
  {NoStop}%
\bibitem [{\citenamefont {Atas}\ \emph {et~al.}(2013)\citenamefont {Atas},
  \citenamefont {Bogomolny}, \citenamefont {Giraud},\ and\ \citenamefont
  {Roux}}]{atas_2013_distribution}%
  \BibitemOpen
  \bibfield  {author} {\bibinfo {author} {\bibfnamefont {Y.~Y.}\ \bibnamefont
  {Atas}}, \bibinfo {author} {\bibfnamefont {E.}~\bibnamefont {Bogomolny}},
  \bibinfo {author} {\bibfnamefont {O.}~\bibnamefont {Giraud}},\ and\ \bibinfo
  {author} {\bibfnamefont {G.}~\bibnamefont {Roux}},\ }\href
  {https://doi.org/10.1103/PhysRevLett.110.084101} {\bibfield  {journal}
  {\bibinfo  {journal} {Phys. Rev. Lett.}\ }\textbf {\bibinfo {volume} {110}},\
  \bibinfo {pages} {084101} (\bibinfo {year} {2013})}\BibitemShut {NoStop}%
\bibitem [{\citenamefont {Mehta}(2004)}]{mehta_2004_random}%
  \BibitemOpen
  \bibfield  {author} {\bibinfo {author} {\bibfnamefont {M.~L.}\ \bibnamefont
  {Mehta}},\ }\href@noop {} {\emph {\bibinfo {title} {Random {{Matrices}}}}}\
  (\bibinfo  {publisher} {Elsevier},\ \bibinfo {year} {2004})\BibitemShut
  {NoStop}%
\bibitem [{\citenamefont {Tekur}\ and\ \citenamefont
  {Santhanam}(2020)}]{tekur_2020_symmetry}%
  \BibitemOpen
  \bibfield  {author} {\bibinfo {author} {\bibfnamefont {S.~H.}\ \bibnamefont
  {Tekur}}\ and\ \bibinfo {author} {\bibfnamefont {M.~S.}\ \bibnamefont
  {Santhanam}},\ }\href {https://doi.org/10.1103/PhysRevResearch.2.032063}
  {\bibfield  {journal} {\bibinfo  {journal} {Phys. Rev. Research}\ }\textbf
  {\bibinfo {volume} {2}},\ \bibinfo {pages} {032063} (\bibinfo {year}
  {2020})}\BibitemShut {NoStop}%
\bibitem [{\citenamefont {Haake}(2001)}]{haake_2001_quantum}%
  \BibitemOpen
  \bibfield  {author} {\bibinfo {author} {\bibfnamefont {F.}~\bibnamefont
  {Haake}},\ }\href@noop {} {\emph {\bibinfo {title} {Quantum {{Signatures}} of
  {{Chaos}}}}}\ (\bibinfo  {publisher} {Springer Science \& Business Media},\
  \bibinfo {year} {2001})\BibitemShut {NoStop}%
\bibitem [{\citenamefont {Roushan}\ \emph {et~al.}(2017)\citenamefont
  {Roushan}, \citenamefont {Neill}, \citenamefont {Tangpanitanon},
  \citenamefont {Bastidas}, \citenamefont {Megrant}, \citenamefont {Barends},
  \citenamefont {Chen}, \citenamefont {Chen}, \citenamefont {Chiaro},
  \citenamefont {Dunsworth}, \citenamefont {Fowler}, \citenamefont {Foxen},
  \citenamefont {Giustina}, \citenamefont {Jeffrey}, \citenamefont {Kelly},
  \citenamefont {Lucero}, \citenamefont {Mutus}, \citenamefont {Neeley},
  \citenamefont {Quintana}, \citenamefont {Sank}, \citenamefont {Vainsencher},
  \citenamefont {Wenner}, \citenamefont {White}, \citenamefont {Neven},
  \citenamefont {Angelakis},\ and\ \citenamefont
  {Martinis}}]{roushan2017spectroscopic}%
  \BibitemOpen
  \bibfield  {author} {\bibinfo {author} {\bibfnamefont {P.}~\bibnamefont
  {Roushan}}, \bibinfo {author} {\bibfnamefont {C.}~\bibnamefont {Neill}},
  \bibinfo {author} {\bibfnamefont {J.}~\bibnamefont {Tangpanitanon}}, \bibinfo
  {author} {\bibfnamefont {V.~M.}\ \bibnamefont {Bastidas}}, \bibinfo {author}
  {\bibfnamefont {A.}~\bibnamefont {Megrant}}, \bibinfo {author} {\bibfnamefont
  {R.}~\bibnamefont {Barends}}, \bibinfo {author} {\bibfnamefont
  {Y.}~\bibnamefont {Chen}}, \bibinfo {author} {\bibfnamefont {Z.}~\bibnamefont
  {Chen}}, \bibinfo {author} {\bibfnamefont {B.}~\bibnamefont {Chiaro}},
  \bibinfo {author} {\bibfnamefont {A.}~\bibnamefont {Dunsworth}}, \bibinfo
  {author} {\bibfnamefont {A.}~\bibnamefont {Fowler}}, \bibinfo {author}
  {\bibfnamefont {B.}~\bibnamefont {Foxen}}, \bibinfo {author} {\bibfnamefont
  {M.}~\bibnamefont {Giustina}}, \bibinfo {author} {\bibfnamefont
  {E.}~\bibnamefont {Jeffrey}}, \bibinfo {author} {\bibfnamefont
  {J.}~\bibnamefont {Kelly}}, \bibinfo {author} {\bibfnamefont
  {E.}~\bibnamefont {Lucero}}, \bibinfo {author} {\bibfnamefont
  {J.}~\bibnamefont {Mutus}}, \bibinfo {author} {\bibfnamefont
  {M.}~\bibnamefont {Neeley}}, \bibinfo {author} {\bibfnamefont
  {C.}~\bibnamefont {Quintana}}, \bibinfo {author} {\bibfnamefont
  {D.}~\bibnamefont {Sank}}, \bibinfo {author} {\bibfnamefont {A.}~\bibnamefont
  {Vainsencher}}, \bibinfo {author} {\bibfnamefont {J.}~\bibnamefont {Wenner}},
  \bibinfo {author} {\bibfnamefont {T.}~\bibnamefont {White}}, \bibinfo
  {author} {\bibfnamefont {H.}~\bibnamefont {Neven}}, \bibinfo {author}
  {\bibfnamefont {D.~G.}\ \bibnamefont {Angelakis}},\ and\ \bibinfo {author}
  {\bibfnamefont {J.}~\bibnamefont {Martinis}},\ }\href
  {https://doi.org/10.1126/science.aao1401} {\bibfield  {journal} {\bibinfo
  {journal} {Science}\ }\textbf {\bibinfo {volume} {358}},\ \bibinfo {pages}
  {1175} (\bibinfo {year} {2017})}\BibitemShut {NoStop}%
\bibitem [{\citenamefont {Dong}\ \emph {et~al.}(2025)\citenamefont {Dong},
  \citenamefont {Zhang}, \citenamefont {Da{\u g}}, \citenamefont {Gao},
  \citenamefont {Wang}, \citenamefont {Deng}, \citenamefont {Zhang},
  \citenamefont {Chen}, \citenamefont {Xu}, \citenamefont {Wang}, \citenamefont
  {Wu}, \citenamefont {Zhang}, \citenamefont {Jin}, \citenamefont {Zhu},
  \citenamefont {Zhang}, \citenamefont {Zou}, \citenamefont {Tan},
  \citenamefont {Cui}, \citenamefont {Zhu}, \citenamefont {Shen}, \citenamefont
  {Li}, \citenamefont {Zhong}, \citenamefont {Bao}, \citenamefont {Li},
  \citenamefont {Wang}, \citenamefont {Guo}, \citenamefont {Song},
  \citenamefont {Liu}, \citenamefont {Chan}, \citenamefont {Ying},\ and\
  \citenamefont {Wang}}]{dong_2025_measuring}%
  \BibitemOpen
  \bibfield  {author} {\bibinfo {author} {\bibfnamefont {H.}~\bibnamefont
  {Dong}}, \bibinfo {author} {\bibfnamefont {P.}~\bibnamefont {Zhang}},
  \bibinfo {author} {\bibfnamefont {C.~B.}\ \bibnamefont {Da{\u g}}}, \bibinfo
  {author} {\bibfnamefont {Y.}~\bibnamefont {Gao}}, \bibinfo {author}
  {\bibfnamefont {N.}~\bibnamefont {Wang}}, \bibinfo {author} {\bibfnamefont
  {J.}~\bibnamefont {Deng}}, \bibinfo {author} {\bibfnamefont {X.}~\bibnamefont
  {Zhang}}, \bibinfo {author} {\bibfnamefont {J.}~\bibnamefont {Chen}},
  \bibinfo {author} {\bibfnamefont {S.}~\bibnamefont {Xu}}, \bibinfo {author}
  {\bibfnamefont {K.}~\bibnamefont {Wang}}, \bibinfo {author} {\bibfnamefont
  {Y.}~\bibnamefont {Wu}}, \bibinfo {author} {\bibfnamefont {C.}~\bibnamefont
  {Zhang}}, \bibinfo {author} {\bibfnamefont {F.}~\bibnamefont {Jin}}, \bibinfo
  {author} {\bibfnamefont {X.}~\bibnamefont {Zhu}}, \bibinfo {author}
  {\bibfnamefont {A.}~\bibnamefont {Zhang}}, \bibinfo {author} {\bibfnamefont
  {Y.}~\bibnamefont {Zou}}, \bibinfo {author} {\bibfnamefont {Z.}~\bibnamefont
  {Tan}}, \bibinfo {author} {\bibfnamefont {Z.}~\bibnamefont {Cui}}, \bibinfo
  {author} {\bibfnamefont {Z.}~\bibnamefont {Zhu}}, \bibinfo {author}
  {\bibfnamefont {F.}~\bibnamefont {Shen}}, \bibinfo {author} {\bibfnamefont
  {T.}~\bibnamefont {Li}}, \bibinfo {author} {\bibfnamefont {J.}~\bibnamefont
  {Zhong}}, \bibinfo {author} {\bibfnamefont {Z.}~\bibnamefont {Bao}}, \bibinfo
  {author} {\bibfnamefont {H.}~\bibnamefont {Li}}, \bibinfo {author}
  {\bibfnamefont {Z.}~\bibnamefont {Wang}}, \bibinfo {author} {\bibfnamefont
  {Q.}~\bibnamefont {Guo}}, \bibinfo {author} {\bibfnamefont {C.}~\bibnamefont
  {Song}}, \bibinfo {author} {\bibfnamefont {F.}~\bibnamefont {Liu}}, \bibinfo
  {author} {\bibfnamefont {A.}~\bibnamefont {Chan}}, \bibinfo {author}
  {\bibfnamefont {L.}~\bibnamefont {Ying}},\ and\ \bibinfo {author}
  {\bibfnamefont {H.}~\bibnamefont {Wang}},\ }\href
  {https://doi.org/10.1103/PhysRevLett.134.010402} {\bibfield  {journal}
  {\bibinfo  {journal} {Phys. Rev. Lett.}\ }\textbf {\bibinfo {volume} {134}},\
  \bibinfo {pages} {010402} (\bibinfo {year} {2025})}\BibitemShut {NoStop}%
\bibitem [{\citenamefont {Mirkin}\ and\ \citenamefont
  {Wisniacki}(2021)}]{mirkin_2021_quantum}%
  \BibitemOpen
  \bibfield  {author} {\bibinfo {author} {\bibfnamefont {N.}~\bibnamefont
  {Mirkin}}\ and\ \bibinfo {author} {\bibfnamefont {D.}~\bibnamefont
  {Wisniacki}},\ }\href {https://doi.org/10.1103/PhysRevE.103.L020201}
  {\bibfield  {journal} {\bibinfo  {journal} {Phys. Rev. E}\ }\textbf {\bibinfo
  {volume} {103}},\ \bibinfo {pages} {L020201} (\bibinfo {year}
  {2021})}\BibitemShut {NoStop}%
\bibitem [{\citenamefont {Mirkin}\ \emph {et~al.}(2021)\citenamefont {Mirkin},
  \citenamefont {Wisniacki}, \citenamefont {Villar},\ and\ \citenamefont
  {Lombardo}}]{mirkin_2021_sensing}%
  \BibitemOpen
  \bibfield  {author} {\bibinfo {author} {\bibfnamefont {N.}~\bibnamefont
  {Mirkin}}, \bibinfo {author} {\bibfnamefont {D.~A.}\ \bibnamefont
  {Wisniacki}}, \bibinfo {author} {\bibfnamefont {P.~I.}\ \bibnamefont
  {Villar}},\ and\ \bibinfo {author} {\bibfnamefont {F.~C.}\ \bibnamefont
  {Lombardo}},\ }\href {https://doi.org/10.1088/2058-9565/ac1e37} {\bibfield
  {journal} {\bibinfo  {journal} {Quantum Sci. Technol.}\ }\textbf {\bibinfo
  {volume} {6}},\ \bibinfo {pages} {045018} (\bibinfo {year}
  {2021})}\BibitemShut {NoStop}%
\bibitem [{\citenamefont {{Vallejo-Fabila}}\ \emph {et~al.}(2025)\citenamefont
  {{Vallejo-Fabila}}, \citenamefont {Das}, \citenamefont {Choudhury},\ and\
  \citenamefont {Santos}}]{vallejo-fabila_2025_singlesite}%
  \BibitemOpen
  \bibfield  {author} {\bibinfo {author} {\bibfnamefont {I.}~\bibnamefont
  {{Vallejo-Fabila}}}, \bibinfo {author} {\bibfnamefont {A.~K.}\ \bibnamefont
  {Das}}, \bibinfo {author} {\bibfnamefont {S.}~\bibnamefont {Choudhury}},\
  and\ \bibinfo {author} {\bibfnamefont {L.~F.}\ \bibnamefont {Santos}},\
  }\href {https://doi.org/10.1103/9tt6-fkf6} {\bibfield  {journal} {\bibinfo
  {journal} {Phys. Rev. E}\ }\textbf {\bibinfo {volume} {112}},\ \bibinfo
  {pages} {044208} (\bibinfo {year} {2025})}\BibitemShut {NoStop}%
\bibitem [{\citenamefont {Breuer}\ and\ \citenamefont
  {Petruccione}(2002)}]{breuer_2002_theory}%
  \BibitemOpen
  \bibfield  {author} {\bibinfo {author} {\bibfnamefont {H.-P.}\ \bibnamefont
  {Breuer}}\ and\ \bibinfo {author} {\bibfnamefont {F.}~\bibnamefont
  {Petruccione}},\ }\href@noop {} {\emph {\bibinfo {title} {The {{Theory}} of
  {{Open Quantum Systems}}}}}\ (\bibinfo  {publisher} {Oxford University
  Press},\ \bibinfo {year} {2002})\BibitemShut {NoStop}%
\bibitem [{\citenamefont {Nielsen}\ and\ \citenamefont
  {Chuang}(2010)}]{nielsen_2010_quantum}%
  \BibitemOpen
  \bibfield  {author} {\bibinfo {author} {\bibfnamefont {M.~A.}\ \bibnamefont
  {Nielsen}}\ and\ \bibinfo {author} {\bibfnamefont {I.~L.}\ \bibnamefont
  {Chuang}},\ }\href@noop {} {\emph {\bibinfo {title} {Quantum {{Computation}}
  and {{Quantum Information}}: 10th {{Anniversary Edition}}}}}\ (\bibinfo
  {publisher} {Cambridge University Press},\ \bibinfo {year}
  {2010})\BibitemShut {NoStop}%
\bibitem [{\citenamefont {Bengtsson}\ and\ \citenamefont
  {Zyczkowski}(2006)}]{bengtsson_2006_geometry}%
  \BibitemOpen
  \bibfield  {author} {\bibinfo {author} {\bibfnamefont {I.}~\bibnamefont
  {Bengtsson}}\ and\ \bibinfo {author} {\bibfnamefont {K.}~\bibnamefont
  {Zyczkowski}},\ }\href {https://doi.org/10.1017/CBO9780511535048} {\emph
  {\bibinfo {title} {Geometry of {{Quantum States}}: {{An Introduction}} to
  {{Quantum Entanglement}}}}}\ (\bibinfo  {publisher} {Cambridge University
  Press},\ \bibinfo {address} {Cambridge},\ \bibinfo {year} {2006})\BibitemShut
  {NoStop}%
\bibitem [{\citenamefont {{\v Z}nidari{\v c}}\ \emph
  {et~al.}(2011)\citenamefont {{\v Z}nidari{\v c}}, \citenamefont {Pineda},\
  and\ \citenamefont {{Garc{\'i}a-Mata}}}]{znidaric_2011_nonmarkovian}%
  \BibitemOpen
  \bibfield  {author} {\bibinfo {author} {\bibfnamefont {M.}~\bibnamefont {{\v
  Z}nidari{\v c}}}, \bibinfo {author} {\bibfnamefont {C.}~\bibnamefont
  {Pineda}},\ and\ \bibinfo {author} {\bibfnamefont {I.}~\bibnamefont
  {{Garc{\'i}a-Mata}}},\ }\href
  {https://doi.org/10.1103/PhysRevLett.107.080404} {\bibfield  {journal}
  {\bibinfo  {journal} {Phys. Rev. Lett.}\ }\textbf {\bibinfo {volume} {107}},\
  \bibinfo {pages} {080404} (\bibinfo {year} {2011})}\BibitemShut {NoStop}%
\bibitem [{\citenamefont {Wallman}\ \emph {et~al.}(2015)\citenamefont
  {Wallman}, \citenamefont {Granade}, \citenamefont {Harper},\ and\
  \citenamefont {Flammia}}]{wallman_2015_estimating}%
  \BibitemOpen
  \bibfield  {author} {\bibinfo {author} {\bibfnamefont {J.}~\bibnamefont
  {Wallman}}, \bibinfo {author} {\bibfnamefont {C.}~\bibnamefont {Granade}},
  \bibinfo {author} {\bibfnamefont {R.}~\bibnamefont {Harper}},\ and\ \bibinfo
  {author} {\bibfnamefont {S.~T.}\ \bibnamefont {Flammia}},\ }\href
  {https://doi.org/10.1088/1367-2630/17/11/113020} {\bibfield  {journal}
  {\bibinfo  {journal} {New J. Phys.}\ }\textbf {\bibinfo {volume} {17}},\
  \bibinfo {pages} {113020} (\bibinfo {year} {2015})}\BibitemShut {NoStop}%
\bibitem [{\citenamefont {Huang}\ \emph {et~al.}(2020)\citenamefont {Huang},
  \citenamefont {Kueng},\ and\ \citenamefont
  {Preskill}}]{huang_2020_predicting}%
  \BibitemOpen
  \bibfield  {author} {\bibinfo {author} {\bibfnamefont {H.-Y.}\ \bibnamefont
  {Huang}}, \bibinfo {author} {\bibfnamefont {R.}~\bibnamefont {Kueng}},\ and\
  \bibinfo {author} {\bibfnamefont {J.}~\bibnamefont {Preskill}},\ }\href
  {https://doi.org/10.1038/s41567-020-0932-7} {\bibfield  {journal} {\bibinfo
  {journal} {Nat. Phys.}\ }\textbf {\bibinfo {volume} {16}},\ \bibinfo {pages}
  {1050} (\bibinfo {year} {2020})}\BibitemShut {NoStop}%
\bibitem [{\citenamefont {Rosgen}(2011)}]{rosgen_2011_testing}%
  \BibitemOpen
  \bibfield  {author} {\bibinfo {author} {\bibfnamefont {B.}~\bibnamefont
  {Rosgen}},\ }in\ \href {https://doi.org/10.1007/978-3-642-18073-6_6} {\emph
  {\bibinfo {booktitle} {Theory of {{Quantum Computation}}, {{Communication}},
  and {{Cryptography}}}}},\ \bibinfo {editor} {edited by\ \bibinfo {editor}
  {\bibfnamefont {W.}~\bibnamefont {{van Dam}}}, \bibinfo {editor}
  {\bibfnamefont {V.~M.}\ \bibnamefont {Kendon}},\ and\ \bibinfo {editor}
  {\bibfnamefont {S.}~\bibnamefont {Severini}}}\ (\bibinfo  {publisher}
  {Springer},\ \bibinfo {address} {Berlin, Heidelberg},\ \bibinfo {year}
  {2011})\ pp.\ \bibinfo {pages} {63--76}\BibitemShut {NoStop}%
\bibitem [{\citenamefont {Yuan}\ \emph {et~al.}(2021)\citenamefont {Yuan},
  \citenamefont {Liu}, \citenamefont {Zhao}, \citenamefont {Regula},
  \citenamefont {Thompson},\ and\ \citenamefont {Gu}}]{yuan_2021_universal}%
  \BibitemOpen
  \bibfield  {author} {\bibinfo {author} {\bibfnamefont {X.}~\bibnamefont
  {Yuan}}, \bibinfo {author} {\bibfnamefont {Y.}~\bibnamefont {Liu}}, \bibinfo
  {author} {\bibfnamefont {Q.}~\bibnamefont {Zhao}}, \bibinfo {author}
  {\bibfnamefont {B.}~\bibnamefont {Regula}}, \bibinfo {author} {\bibfnamefont
  {J.}~\bibnamefont {Thompson}},\ and\ \bibinfo {author} {\bibfnamefont
  {M.}~\bibnamefont {Gu}},\ }\href {https://doi.org/10.1038/s41534-021-00444-9}
  {\bibfield  {journal} {\bibinfo  {journal} {npj Quantum Inf.}\ }\textbf
  {\bibinfo {volume} {7}},\ \bibinfo {pages} {108} (\bibinfo {year}
  {2021})}\BibitemShut {NoStop}%
\bibitem [{\citenamefont {Islam}\ \emph {et~al.}(2015)\citenamefont {Islam},
  \citenamefont {Ma}, \citenamefont {Preiss}, \citenamefont {Eric~Tai},
  \citenamefont {Lukin}, \citenamefont {Rispoli},\ and\ \citenamefont
  {Greiner}}]{islam_2015_measuring}%
  \BibitemOpen
  \bibfield  {author} {\bibinfo {author} {\bibfnamefont {R.}~\bibnamefont
  {Islam}}, \bibinfo {author} {\bibfnamefont {R.}~\bibnamefont {Ma}}, \bibinfo
  {author} {\bibfnamefont {P.~M.}\ \bibnamefont {Preiss}}, \bibinfo {author}
  {\bibfnamefont {M.}~\bibnamefont {Eric~Tai}}, \bibinfo {author}
  {\bibfnamefont {A.}~\bibnamefont {Lukin}}, \bibinfo {author} {\bibfnamefont
  {M.}~\bibnamefont {Rispoli}},\ and\ \bibinfo {author} {\bibfnamefont
  {M.}~\bibnamefont {Greiner}},\ }\href {https://doi.org/10.1038/nature15750}
  {\bibfield  {journal} {\bibinfo  {journal} {Nature}\ }\textbf {\bibinfo
  {volume} {528}},\ \bibinfo {pages} {77} (\bibinfo {year} {2015})}\BibitemShut
  {NoStop}%
\bibitem [{\citenamefont {Kaufman}\ \emph {et~al.}(2016)\citenamefont
  {Kaufman}, \citenamefont {Tai}, \citenamefont {Lukin}, \citenamefont
  {Rispoli}, \citenamefont {Schittko}, \citenamefont {Preiss},\ and\
  \citenamefont {Greiner}}]{kaufman_2016_quantum}%
  \BibitemOpen
  \bibfield  {author} {\bibinfo {author} {\bibfnamefont {A.~M.}\ \bibnamefont
  {Kaufman}}, \bibinfo {author} {\bibfnamefont {M.~E.}\ \bibnamefont {Tai}},
  \bibinfo {author} {\bibfnamefont {A.}~\bibnamefont {Lukin}}, \bibinfo
  {author} {\bibfnamefont {M.}~\bibnamefont {Rispoli}}, \bibinfo {author}
  {\bibfnamefont {R.}~\bibnamefont {Schittko}}, \bibinfo {author}
  {\bibfnamefont {P.~M.}\ \bibnamefont {Preiss}},\ and\ \bibinfo {author}
  {\bibfnamefont {M.}~\bibnamefont {Greiner}},\ }\href
  {https://doi.org/10.1126/science.aaf6725} {\bibfield  {journal} {\bibinfo
  {journal} {Science}\ }\textbf {\bibinfo {volume} {353}},\ \bibinfo {pages}
  {794} (\bibinfo {year} {2016})}\BibitemShut {NoStop}%
\bibitem [{\citenamefont {Jamio{\l}kowski}(1972)}]{jamiolkowski_1972_linear}%
  \BibitemOpen
  \bibfield  {author} {\bibinfo {author} {\bibfnamefont {A.}~\bibnamefont
  {Jamio{\l}kowski}},\ }\href {https://doi.org/10.1016/0034-4877(72)90011-0}
  {\bibfield  {journal} {\bibinfo  {journal} {Rep. Math. Phys.}\ }\textbf
  {\bibinfo {volume} {3}},\ \bibinfo {pages} {275} (\bibinfo {year}
  {1972})}\BibitemShut {NoStop}%
\bibitem [{\citenamefont {Choi}(1975)}]{choi_1975_completely}%
  \BibitemOpen
  \bibfield  {author} {\bibinfo {author} {\bibfnamefont {M.-D.}\ \bibnamefont
  {Choi}},\ }\href {https://doi.org/10.1016/0024-3795(75)90075-0} {\bibfield
  {journal} {\bibinfo  {journal} {Linear Algebra Appl.}\ }\textbf {\bibinfo
  {volume} {10}},\ \bibinfo {pages} {285} (\bibinfo {year} {1975})}\BibitemShut
  {NoStop}%
\bibitem [{\citenamefont {Gour}\ and\ \citenamefont
  {Wilde}(2021)}]{gour_2021_entropy}%
  \BibitemOpen
  \bibfield  {author} {\bibinfo {author} {\bibfnamefont {G.}~\bibnamefont
  {Gour}}\ and\ \bibinfo {author} {\bibfnamefont {M.~M.}\ \bibnamefont
  {Wilde}},\ }\href {https://doi.org/10.1103/PhysRevResearch.3.023096}
  {\bibfield  {journal} {\bibinfo  {journal} {Phys. Rev. Research}\ }\textbf
  {\bibinfo {volume} {3}},\ \bibinfo {pages} {023096} (\bibinfo {year}
  {2021})}\BibitemShut {NoStop}%
\bibitem [{\citenamefont {Chu}\ \emph {et~al.}(2022)\citenamefont {Chu},
  \citenamefont {Huang}, \citenamefont {Li},\ and\ \citenamefont
  {Zheng}}]{chu_2022_entropy}%
  \BibitemOpen
  \bibfield  {author} {\bibinfo {author} {\bibfnamefont {Y.}~\bibnamefont
  {Chu}}, \bibinfo {author} {\bibfnamefont {F.}~\bibnamefont {Huang}}, \bibinfo
  {author} {\bibfnamefont {M.-X.}\ \bibnamefont {Li}},\ and\ \bibinfo {author}
  {\bibfnamefont {Z.-J.}\ \bibnamefont {Zheng}},\ }\href
  {https://doi.org/10.1007/s11128-022-03778-1} {\bibfield  {journal} {\bibinfo
  {journal} {Quantum Inf. Process.}\ }\textbf {\bibinfo {volume} {22}},\
  \bibinfo {pages} {27} (\bibinfo {year} {2022})}\BibitemShut {NoStop}%
\bibitem [{\citenamefont {Li}\ \emph {et~al.}(2025)\citenamefont {Li},
  \citenamefont {Luo}, \citenamefont {Sun},\ and\ \citenamefont
  {Wang}}]{li_2025_quantifying}%
  \BibitemOpen
  \bibfield  {author} {\bibinfo {author} {\bibfnamefont {Y.}~\bibnamefont
  {Li}}, \bibinfo {author} {\bibfnamefont {S.}~\bibnamefont {Luo}}, \bibinfo
  {author} {\bibfnamefont {Y.}~\bibnamefont {Sun}},\ and\ \bibinfo {author}
  {\bibfnamefont {S.}~\bibnamefont {Wang}},\ }\href
  {https://doi.org/10.1103/ckjg-xpmz} {\bibfield  {journal} {\bibinfo
  {journal} {Phys. Rev. A}\ }\textbf {\bibinfo {volume} {112}},\ \bibinfo
  {pages} {062440} (\bibinfo {year} {2025})}\BibitemShut {NoStop}%
\bibitem [{\citenamefont {Roga}\ \emph {et~al.}(2011)\citenamefont {Roga},
  \citenamefont {{\.Z}yczkowski},\ and\ \citenamefont
  {Fannes}}]{roga_2011_entropic}%
  \BibitemOpen
  \bibfield  {author} {\bibinfo {author} {\bibfnamefont {W.}~\bibnamefont
  {Roga}}, \bibinfo {author} {\bibfnamefont {K.}~\bibnamefont
  {{\.Z}yczkowski}},\ and\ \bibinfo {author} {\bibfnamefont {M.}~\bibnamefont
  {Fannes}},\ }\href {https://doi.org/10.1142/S0219749911007794} {\bibfield
  {journal} {\bibinfo  {journal} {Int. J. Quantum Inf.}\ }\textbf {\bibinfo
  {volume} {09}},\ \bibinfo {pages} {1031} (\bibinfo {year}
  {2011})}\BibitemShut {NoStop}%
\bibitem [{\citenamefont {Korzekwa}\ \emph {et~al.}(2018)\citenamefont
  {Korzekwa}, \citenamefont {Czach{\'o}rski}, \citenamefont {Pucha{\l}a},\ and\
  \citenamefont {{\.Z}yczkowski}}]{korzekwa_2018_coherifying}%
  \BibitemOpen
  \bibfield  {author} {\bibinfo {author} {\bibfnamefont {K.}~\bibnamefont
  {Korzekwa}}, \bibinfo {author} {\bibfnamefont {S.}~\bibnamefont
  {Czach{\'o}rski}}, \bibinfo {author} {\bibfnamefont {Z.}~\bibnamefont
  {Pucha{\l}a}},\ and\ \bibinfo {author} {\bibfnamefont {K.}~\bibnamefont
  {{\.Z}yczkowski}},\ }\href {https://doi.org/10.1088/1367-2630/aaaff3}
  {\bibfield  {journal} {\bibinfo  {journal} {New J. Phys.}\ }\textbf {\bibinfo
  {volume} {20}},\ \bibinfo {pages} {043028} (\bibinfo {year}
  {2018})}\BibitemShut {NoStop}%
\bibitem [{\citenamefont {Mele}(2024)}]{mele_2024_introduction}%
  \BibitemOpen
  \bibfield  {author} {\bibinfo {author} {\bibfnamefont {A.~A.}\ \bibnamefont
  {Mele}},\ }\href {https://doi.org/10.22331/q-2024-05-08-1340} {\bibfield
  {journal} {\bibinfo  {journal} {Quantum}\ }\textbf {\bibinfo {volume} {8}},\
  \bibinfo {pages} {1340} (\bibinfo {year} {2024})}\BibitemShut {NoStop}%
\bibitem [{\citenamefont {Peres}(1984)}]{peres_1984_stability}%
  \BibitemOpen
  \bibfield  {author} {\bibinfo {author} {\bibfnamefont {A.}~\bibnamefont
  {Peres}},\ }\href {https://doi.org/10.1103/PhysRevA.30.1610} {\bibfield
  {journal} {\bibinfo  {journal} {Phys. Rev. A}\ }\textbf {\bibinfo {volume}
  {30}},\ \bibinfo {pages} {1610} (\bibinfo {year} {1984})}\BibitemShut
  {NoStop}%
\bibitem [{\citenamefont {Gorin}\ \emph {et~al.}(2006)\citenamefont {Gorin},
  \citenamefont {Prosen}, \citenamefont {Seligman},\ and\ \citenamefont {{\v
  Z}nidari{\v c}}}]{gorin_2006_dynamics}%
  \BibitemOpen
  \bibfield  {author} {\bibinfo {author} {\bibfnamefont {T.}~\bibnamefont
  {Gorin}}, \bibinfo {author} {\bibfnamefont {T.}~\bibnamefont {Prosen}},
  \bibinfo {author} {\bibfnamefont {T.~H.}\ \bibnamefont {Seligman}},\ and\
  \bibinfo {author} {\bibfnamefont {M.}~\bibnamefont {{\v Z}nidari{\v c}}},\
  }\href {https://doi.org/10.1016/j.physrep.2006.09.003} {\bibfield  {journal}
  {\bibinfo  {journal} {Phys. Rep.}\ }\textbf {\bibinfo {volume} {435}},\
  \bibinfo {pages} {33} (\bibinfo {year} {2006})}\BibitemShut {NoStop}%
\bibitem [{\citenamefont {Deutsch}(1991)}]{deutsch_1991_quantum}%
  \BibitemOpen
  \bibfield  {author} {\bibinfo {author} {\bibfnamefont {J.~M.}\ \bibnamefont
  {Deutsch}},\ }\href {https://doi.org/10.1103/PhysRevA.43.2046} {\bibfield
  {journal} {\bibinfo  {journal} {Phys. Rev. A}\ }\textbf {\bibinfo {volume}
  {43}},\ \bibinfo {pages} {2046} (\bibinfo {year} {1991})}\BibitemShut
  {NoStop}%
\bibitem [{\citenamefont {Bardarson}\ \emph {et~al.}(2012)\citenamefont
  {Bardarson}, \citenamefont {Pollmann},\ and\ \citenamefont
  {Moore}}]{bardarson2012unbounded}%
  \BibitemOpen
  \bibfield  {author} {\bibinfo {author} {\bibfnamefont {J.~H.}\ \bibnamefont
  {Bardarson}}, \bibinfo {author} {\bibfnamefont {F.}~\bibnamefont
  {Pollmann}},\ and\ \bibinfo {author} {\bibfnamefont {J.~E.}\ \bibnamefont
  {Moore}},\ }\href {https://doi.org/10.1103/PhysRevLett.109.017202} {\bibfield
   {journal} {\bibinfo  {journal} {Phys. Rev. Lett.}\ }\textbf {\bibinfo
  {volume} {109}},\ \bibinfo {pages} {017202} (\bibinfo {year}
  {2012})}\BibitemShut {NoStop}%
\bibitem [{\citenamefont {Santos}\ and\ \citenamefont
  {Rigol}(2010)}]{santos2010onset}%
  \BibitemOpen
  \bibfield  {author} {\bibinfo {author} {\bibfnamefont {L.~F.}\ \bibnamefont
  {Santos}}\ and\ \bibinfo {author} {\bibfnamefont {M.}~\bibnamefont {Rigol}},\
  }\href {https://doi.org/10.1103/PhysRevE.81.036206} {\bibfield  {journal}
  {\bibinfo  {journal} {Phys. Rev. E}\ }\textbf {\bibinfo {volume} {81}},\
  \bibinfo {pages} {036206} (\bibinfo {year} {2010})}\BibitemShut {NoStop}%
\bibitem [{\citenamefont {Alba}\ and\ \citenamefont
  {Calabrese}(2017)}]{alba2017entanglement}%
  \BibitemOpen
  \bibfield  {author} {\bibinfo {author} {\bibfnamefont {V.}~\bibnamefont
  {Alba}}\ and\ \bibinfo {author} {\bibfnamefont {P.}~\bibnamefont
  {Calabrese}},\ }\href {https://doi.org/10.1073/pnas.1703516114} {\bibfield
  {journal} {\bibinfo  {journal} {Proc. Natl. Acad. Sci. U.S.A.}\ }\textbf
  {\bibinfo {volume} {114}},\ \bibinfo {pages} {7947} (\bibinfo {year}
  {2017})}\BibitemShut {NoStop}%
\bibitem [{\citenamefont {Smith}\ \emph {et~al.}(2016)\citenamefont {Smith},
  \citenamefont {Lee}, \citenamefont {Richerme}, \citenamefont {Neyenhuis},
  \citenamefont {Hess}, \citenamefont {Hauke}, \citenamefont {Heyl},
  \citenamefont {Huse},\ and\ \citenamefont {Monroe}}]{smith2016many}%
  \BibitemOpen
  \bibfield  {author} {\bibinfo {author} {\bibfnamefont {J.}~\bibnamefont
  {Smith}}, \bibinfo {author} {\bibfnamefont {A.}~\bibnamefont {Lee}}, \bibinfo
  {author} {\bibfnamefont {P.}~\bibnamefont {Richerme}}, \bibinfo {author}
  {\bibfnamefont {B.}~\bibnamefont {Neyenhuis}}, \bibinfo {author}
  {\bibfnamefont {P.~W.}\ \bibnamefont {Hess}}, \bibinfo {author}
  {\bibfnamefont {P.}~\bibnamefont {Hauke}}, \bibinfo {author} {\bibfnamefont
  {M.}~\bibnamefont {Heyl}}, \bibinfo {author} {\bibfnamefont {D.~A.}\
  \bibnamefont {Huse}},\ and\ \bibinfo {author} {\bibfnamefont
  {C.}~\bibnamefont {Monroe}},\ }\href {https://doi.org/10.1038/nphys3783}
  {\bibfield  {journal} {\bibinfo  {journal} {Nat. Phys.}\ }\textbf {\bibinfo
  {volume} {12}},\ \bibinfo {pages} {907} (\bibinfo {year} {2016})}\BibitemShut
  {NoStop}%
\bibitem [{\citenamefont {Berry}\ and\ \citenamefont
  {Tabor}(1977)}]{berry_1977_level}%
  \BibitemOpen
  \bibfield  {author} {\bibinfo {author} {\bibfnamefont {M.~V.}\ \bibnamefont
  {Berry}}\ and\ \bibinfo {author} {\bibfnamefont {M.}~\bibnamefont {Tabor}},\
  }\href {https://doi.org/10.1098/rspa.1977.0140} {\bibfield  {journal}
  {\bibinfo  {journal} {Proc. R. Soc. London, Ser. A}\ }\textbf {\bibinfo
  {volume} {356}},\ \bibinfo {pages} {375} (\bibinfo {year}
  {1977})}\BibitemShut {NoStop}%
\bibitem [{Note1()}]{Note1}%
  \BibitemOpen
  \bibinfo {note} {A rigorous definition of many-body quantum integrability
  requires the thermodynamic limit~\cite {caux_2011_remarks}. For instance, the
  mixed-field Ising model used in this work has been formally shown to be
  non-integrable for any non-zero $J$~\cite {chiba_2024_proof}.}\BibitemShut
  {Stop}%
\bibitem [{\citenamefont {Scialchi}\ \emph {et~al.}(2024)\citenamefont
  {Scialchi}, \citenamefont {Roncaglia},\ and\ \citenamefont
  {Wisniacki}}]{scialchi_2024_integrabilitytochaos}%
  \BibitemOpen
  \bibfield  {author} {\bibinfo {author} {\bibfnamefont {G.~F.}\ \bibnamefont
  {Scialchi}}, \bibinfo {author} {\bibfnamefont {A.~J.}\ \bibnamefont
  {Roncaglia}},\ and\ \bibinfo {author} {\bibfnamefont {D.~A.}\ \bibnamefont
  {Wisniacki}},\ }\href {https://doi.org/10.1103/PhysRevE.109.054209}
  {\bibfield  {journal} {\bibinfo  {journal} {Phys. Rev. E}\ }\textbf {\bibinfo
  {volume} {109}},\ \bibinfo {pages} {054209} (\bibinfo {year}
  {2024})}\BibitemShut {NoStop}%
\bibitem [{\citenamefont {Bernien}\ \emph {et~al.}(2017)\citenamefont
  {Bernien}, \citenamefont {Schwartz}, \citenamefont {Keesling}, \citenamefont
  {Levine}, \citenamefont {Omran}, \citenamefont {Pichler}, \citenamefont
  {Choi}, \citenamefont {Zibrov}, \citenamefont {Endres}, \citenamefont
  {Greiner}, \citenamefont {Vuleti{\'c}},\ and\ \citenamefont
  {Lukin}}]{bernien2017probing}%
  \BibitemOpen
  \bibfield  {author} {\bibinfo {author} {\bibfnamefont {H.}~\bibnamefont
  {Bernien}}, \bibinfo {author} {\bibfnamefont {S.}~\bibnamefont {Schwartz}},
  \bibinfo {author} {\bibfnamefont {A.}~\bibnamefont {Keesling}}, \bibinfo
  {author} {\bibfnamefont {H.}~\bibnamefont {Levine}}, \bibinfo {author}
  {\bibfnamefont {A.}~\bibnamefont {Omran}}, \bibinfo {author} {\bibfnamefont
  {H.}~\bibnamefont {Pichler}}, \bibinfo {author} {\bibfnamefont
  {S.}~\bibnamefont {Choi}}, \bibinfo {author} {\bibfnamefont {A.~S.}\
  \bibnamefont {Zibrov}}, \bibinfo {author} {\bibfnamefont {M.}~\bibnamefont
  {Endres}}, \bibinfo {author} {\bibfnamefont {M.}~\bibnamefont {Greiner}},
  \bibinfo {author} {\bibfnamefont {V.}~\bibnamefont {Vuleti{\'c}}},\ and\
  \bibinfo {author} {\bibfnamefont {M.~D.}\ \bibnamefont {Lukin}},\ }\href
  {https://doi.org/10.1038/nature24622} {\bibfield  {journal} {\bibinfo
  {journal} {Nature}\ }\textbf {\bibinfo {volume} {551}},\ \bibinfo {pages}
  {579} (\bibinfo {year} {2017})}\BibitemShut {NoStop}%
\bibitem [{\citenamefont {Brydges}\ \emph {et~al.}(2019)\citenamefont
  {Brydges}, \citenamefont {Elben}, \citenamefont {Jurcevic}, \citenamefont
  {Vermersch}, \citenamefont {Maier}, \citenamefont {Lanyon}, \citenamefont
  {Zoller}, \citenamefont {Blatt},\ and\ \citenamefont
  {Roos}}]{brydges2019probing}%
  \BibitemOpen
  \bibfield  {author} {\bibinfo {author} {\bibfnamefont {T.}~\bibnamefont
  {Brydges}}, \bibinfo {author} {\bibfnamefont {A.}~\bibnamefont {Elben}},
  \bibinfo {author} {\bibfnamefont {P.}~\bibnamefont {Jurcevic}}, \bibinfo
  {author} {\bibfnamefont {B.}~\bibnamefont {Vermersch}}, \bibinfo {author}
  {\bibfnamefont {C.}~\bibnamefont {Maier}}, \bibinfo {author} {\bibfnamefont
  {B.~P.}\ \bibnamefont {Lanyon}}, \bibinfo {author} {\bibfnamefont
  {P.}~\bibnamefont {Zoller}}, \bibinfo {author} {\bibfnamefont
  {R.}~\bibnamefont {Blatt}},\ and\ \bibinfo {author} {\bibfnamefont {C.~F.}\
  \bibnamefont {Roos}},\ }\href {https://doi.org/10.1126/science.aau4963}
  {\bibfield  {journal} {\bibinfo  {journal} {Science}\ }\textbf {\bibinfo
  {volume} {364}},\ \bibinfo {pages} {260} (\bibinfo {year}
  {2019})}\BibitemShut {NoStop}%
\bibitem [{\citenamefont {Goldstein}\ \emph {et~al.}(2006)\citenamefont
  {Goldstein}, \citenamefont {Lebowitz}, \citenamefont {Tumulka},\ and\
  \citenamefont {Zangh{\`i}}}]{goldstein_2006_canonical}%
  \BibitemOpen
  \bibfield  {author} {\bibinfo {author} {\bibfnamefont {S.}~\bibnamefont
  {Goldstein}}, \bibinfo {author} {\bibfnamefont {J.~L.}\ \bibnamefont
  {Lebowitz}}, \bibinfo {author} {\bibfnamefont {R.}~\bibnamefont {Tumulka}},\
  and\ \bibinfo {author} {\bibfnamefont {N.}~\bibnamefont {Zangh{\`i}}},\
  }\href {https://doi.org/10.1103/PhysRevLett.96.050403} {\bibfield  {journal}
  {\bibinfo  {journal} {Phys. Rev. Lett.}\ }\textbf {\bibinfo {volume} {96}},\
  \bibinfo {pages} {050403} (\bibinfo {year} {2006})}\BibitemShut {NoStop}%
\bibitem [{\citenamefont {Rigol}\ \emph {et~al.}(2008)\citenamefont {Rigol},
  \citenamefont {Dunjko},\ and\ \citenamefont
  {Olshanii}}]{rigol_2008_thermalization}%
  \BibitemOpen
  \bibfield  {author} {\bibinfo {author} {\bibfnamefont {M.}~\bibnamefont
  {Rigol}}, \bibinfo {author} {\bibfnamefont {V.}~\bibnamefont {Dunjko}},\ and\
  \bibinfo {author} {\bibfnamefont {M.}~\bibnamefont {Olshanii}},\ }\href
  {https://doi.org/10.1038/nature06838} {\bibfield  {journal} {\bibinfo
  {journal} {Nature}\ }\textbf {\bibinfo {volume} {452}},\ \bibinfo {pages}
  {854} (\bibinfo {year} {2008})}\BibitemShut {NoStop}%
\bibitem [{Note2()}]{Note2}%
  \BibitemOpen
  \bibinfo {note} {When averaged over the Haar measure, a single-spin state
  $\dyad {\psi _j}$ becomes the maximally mixed state $\protect \mathbb {I}/2$.
  The full environment state $\rho _E$, being a tensor product of $L-1$ such
  states, thus averages to $\langle \rho _E \rangle = \protect \mathbb {I}_E /
  2^{L-1}$, which is the state of infinite-temperature
  equilibrium.}\BibitemShut {Stop}%
\bibitem [{\citenamefont {Sachdev}(2011)}]{sachdev_2011_quantum}%
  \BibitemOpen
  \bibfield  {author} {\bibinfo {author} {\bibfnamefont {S.}~\bibnamefont
  {Sachdev}},\ }\href {https://doi.org/10.1017/CBO9780511973765} {\emph
  {\bibinfo {title} {Quantum {{Phase Transitions}}}}},\ \bibinfo {edition}
  {2nd}\ ed.\ (\bibinfo  {publisher} {Cambridge University Press},\ \bibinfo
  {address} {Cambridge},\ \bibinfo {year} {2011})\BibitemShut {NoStop}%
\bibitem [{\citenamefont {Baxter}(2007)}]{baxter_2007_exactly}%
  \BibitemOpen
  \bibfield  {author} {\bibinfo {author} {\bibfnamefont {R.~J.}\ \bibnamefont
  {Baxter}},\ }\href@noop {} {\emph {\bibinfo {title} {Exactly {{Solved
  Models}} in {{Statistical Mechanics}}}}}\ (\bibinfo  {publisher} {Courier
  Corporation},\ \bibinfo {year} {2007})\BibitemShut {NoStop}%
\bibitem [{\citenamefont {Kim}\ \emph {et~al.}(2014)\citenamefont {Kim},
  \citenamefont {Ikeda},\ and\ \citenamefont {Huse}}]{kim_2014_testing}%
  \BibitemOpen
  \bibfield  {author} {\bibinfo {author} {\bibfnamefont {H.}~\bibnamefont
  {Kim}}, \bibinfo {author} {\bibfnamefont {T.~N.}\ \bibnamefont {Ikeda}},\
  and\ \bibinfo {author} {\bibfnamefont {D.~A.}\ \bibnamefont {Huse}},\ }\href
  {https://doi.org/10.1103/PhysRevE.90.052105} {\bibfield  {journal} {\bibinfo
  {journal} {Phys. Rev. E}\ }\textbf {\bibinfo {volume} {90}},\ \bibinfo
  {pages} {052105} (\bibinfo {year} {2014})}\BibitemShut {NoStop}%
\bibitem [{\citenamefont {Garrison}\ and\ \citenamefont
  {Grover}(2018)}]{garrison_2018_does}%
  \BibitemOpen
  \bibfield  {author} {\bibinfo {author} {\bibfnamefont {J.~R.}\ \bibnamefont
  {Garrison}}\ and\ \bibinfo {author} {\bibfnamefont {T.}~\bibnamefont
  {Grover}},\ }\href {https://doi.org/10.1103/PhysRevX.8.021026} {\bibfield
  {journal} {\bibinfo  {journal} {Phys. Rev. X}\ }\textbf {\bibinfo {volume}
  {8}},\ \bibinfo {pages} {021026} (\bibinfo {year} {2018})}\BibitemShut
  {NoStop}%
\bibitem [{\citenamefont {Borgonovi}\ \emph {et~al.}(2016)\citenamefont
  {Borgonovi}, \citenamefont {Izrailev}, \citenamefont {Santos},\ and\
  \citenamefont {Zelevinsky}}]{borgonovi_2016_quantum}%
  \BibitemOpen
  \bibfield  {author} {\bibinfo {author} {\bibfnamefont {F.}~\bibnamefont
  {Borgonovi}}, \bibinfo {author} {\bibfnamefont {F.~M.}\ \bibnamefont
  {Izrailev}}, \bibinfo {author} {\bibfnamefont {L.~F.}\ \bibnamefont
  {Santos}},\ and\ \bibinfo {author} {\bibfnamefont {V.~G.}\ \bibnamefont
  {Zelevinsky}},\ }\href {https://doi.org/10.1016/j.physrep.2016.02.005}
  {\bibfield  {journal} {\bibinfo  {journal} {Phys. Rep.}\ }\textbf {\bibinfo
  {volume} {626}},\ \bibinfo {pages} {1} (\bibinfo {year} {2016})}\BibitemShut
  {NoStop}%
\bibitem [{\citenamefont {Chiba}(2024)}]{chiba_2024_proof}%
  \BibitemOpen
  \bibfield  {author} {\bibinfo {author} {\bibfnamefont {Y.}~\bibnamefont
  {Chiba}},\ }\href {https://doi.org/10.1103/PhysRevB.109.035123} {\bibfield
  {journal} {\bibinfo  {journal} {Phys. Rev. B}\ }\textbf {\bibinfo {volume}
  {109}},\ \bibinfo {pages} {035123} (\bibinfo {year} {2024})}\BibitemShut
  {NoStop}%
\bibitem [{\citenamefont {Karthik}\ \emph {et~al.}(2007)\citenamefont
  {Karthik}, \citenamefont {Sharma},\ and\ \citenamefont
  {Lakshminarayan}}]{karthik_2007_entanglement}%
  \BibitemOpen
  \bibfield  {author} {\bibinfo {author} {\bibfnamefont {J.}~\bibnamefont
  {Karthik}}, \bibinfo {author} {\bibfnamefont {A.}~\bibnamefont {Sharma}},\
  and\ \bibinfo {author} {\bibfnamefont {A.}~\bibnamefont {Lakshminarayan}},\
  }\href {https://doi.org/10.1103/PhysRevA.75.022304} {\bibfield  {journal}
  {\bibinfo  {journal} {Phys. Rev. A}\ }\textbf {\bibinfo {volume} {75}},\
  \bibinfo {pages} {022304} (\bibinfo {year} {2007})}\BibitemShut {NoStop}%
\bibitem [{\citenamefont {Atas}\ and\ \citenamefont
  {Bogomolny}(2017)}]{atas_2017_quantum}%
  \BibitemOpen
  \bibfield  {author} {\bibinfo {author} {\bibfnamefont {Y.~Y.}\ \bibnamefont
  {Atas}}\ and\ \bibinfo {author} {\bibfnamefont {E.}~\bibnamefont
  {Bogomolny}},\ }\href {https://doi.org/10.1088/1751-8121/aa81f6} {\bibfield
  {journal} {\bibinfo  {journal} {J. Phys. A: Math. Theor.}\ }\textbf {\bibinfo
  {volume} {50}},\ \bibinfo {pages} {385102} (\bibinfo {year}
  {2017})}\BibitemShut {NoStop}%
\bibitem [{\citenamefont {Camargo}\ \emph {et~al.}(2024)\citenamefont
  {Camargo}, \citenamefont {Huh}, \citenamefont {Jahnke}, \citenamefont
  {Jeong}, \citenamefont {Kim},\ and\ \citenamefont
  {Nishida}}]{camargo_2024_spread}%
  \BibitemOpen
  \bibfield  {author} {\bibinfo {author} {\bibfnamefont {H.~A.}\ \bibnamefont
  {Camargo}}, \bibinfo {author} {\bibfnamefont {K.-B.}\ \bibnamefont {Huh}},
  \bibinfo {author} {\bibfnamefont {V.}~\bibnamefont {Jahnke}}, \bibinfo
  {author} {\bibfnamefont {H.-S.}\ \bibnamefont {Jeong}}, \bibinfo {author}
  {\bibfnamefont {K.-Y.}\ \bibnamefont {Kim}},\ and\ \bibinfo {author}
  {\bibfnamefont {M.}~\bibnamefont {Nishida}},\ }\href
  {https://doi.org/10.1007/JHEP08(2024)241} {\bibfield  {journal} {\bibinfo
  {journal} {J. High Energy Phys.}\ }\textbf {\bibinfo {volume} {2024}}\bibinfo
   {number} { (8)},\ \bibinfo {pages} {241}}\BibitemShut {NoStop}%
\bibitem [{\citenamefont {Bethe}(1931)}]{bethe_1931_zur}%
  \BibitemOpen
\bibfield  {number} {  }\bibfield  {author} {\bibinfo {author} {\bibfnamefont
  {H.}~\bibnamefont {Bethe}},\ }\href {https://doi.org/10.1007/BF01341708}
  {\bibfield  {journal} {\bibinfo  {journal} {Z. Phys.}\ }\textbf {\bibinfo
  {volume} {71}},\ \bibinfo {pages} {205} (\bibinfo {year} {1931})}\BibitemShut
  {NoStop}%
\bibitem [{\citenamefont {Luitz}\ \emph {et~al.}(2015)\citenamefont {Luitz},
  \citenamefont {Laflorencie},\ and\ \citenamefont
  {Alet}}]{luitz_2015_manybody}%
  \BibitemOpen
  \bibfield  {author} {\bibinfo {author} {\bibfnamefont {D.~J.}\ \bibnamefont
  {Luitz}}, \bibinfo {author} {\bibfnamefont {N.}~\bibnamefont {Laflorencie}},\
  and\ \bibinfo {author} {\bibfnamefont {F.}~\bibnamefont {Alet}},\ }\href
  {https://doi.org/10.1103/PhysRevB.91.081103} {\bibfield  {journal} {\bibinfo
  {journal} {Phys. Rev. B}\ }\textbf {\bibinfo {volume} {91}},\ \bibinfo
  {pages} {081103} (\bibinfo {year} {2015})}\BibitemShut {NoStop}%
\bibitem [{\citenamefont {Abanin}\ \emph {et~al.}(2019)\citenamefont {Abanin},
  \citenamefont {Altman}, \citenamefont {Bloch},\ and\ \citenamefont
  {Serbyn}}]{abanin_2019_colloquium}%
  \BibitemOpen
  \bibfield  {author} {\bibinfo {author} {\bibfnamefont {D.~A.}\ \bibnamefont
  {Abanin}}, \bibinfo {author} {\bibfnamefont {E.}~\bibnamefont {Altman}},
  \bibinfo {author} {\bibfnamefont {I.}~\bibnamefont {Bloch}},\ and\ \bibinfo
  {author} {\bibfnamefont {M.}~\bibnamefont {Serbyn}},\ }\href
  {https://doi.org/10.1103/RevModPhys.91.021001} {\bibfield  {journal}
  {\bibinfo  {journal} {Rev. Mod. Phys.}\ }\textbf {\bibinfo {volume} {91}},\
  \bibinfo {pages} {021001} (\bibinfo {year} {2019})}\BibitemShut {NoStop}%
\bibitem [{\citenamefont {Santos}(2004)}]{santos_2004_integrability}%
  \BibitemOpen
  \bibfield  {author} {\bibinfo {author} {\bibfnamefont {L.~F.}\ \bibnamefont
  {Santos}},\ }\href {https://doi.org/10.1088/0305-4470/37/17/004} {\bibfield
  {journal} {\bibinfo  {journal} {J. Phys. A: Math. Gen.}\ }\textbf {\bibinfo
  {volume} {37}},\ \bibinfo {pages} {4723} (\bibinfo {year}
  {2004})}\BibitemShut {NoStop}%
\bibitem [{\citenamefont {Brenes}\ \emph {et~al.}(2018)\citenamefont {Brenes},
  \citenamefont {Mascarenhas}, \citenamefont {Rigol},\ and\ \citenamefont
  {Goold}}]{brenes_2018_hightemperature}%
  \BibitemOpen
  \bibfield  {author} {\bibinfo {author} {\bibfnamefont {M.}~\bibnamefont
  {Brenes}}, \bibinfo {author} {\bibfnamefont {E.}~\bibnamefont {Mascarenhas}},
  \bibinfo {author} {\bibfnamefont {M.}~\bibnamefont {Rigol}},\ and\ \bibinfo
  {author} {\bibfnamefont {J.}~\bibnamefont {Goold}},\ }\href
  {https://doi.org/10.1103/PhysRevB.98.235128} {\bibfield  {journal} {\bibinfo
  {journal} {Phys. Rev. B}\ }\textbf {\bibinfo {volume} {98}},\ \bibinfo
  {pages} {235128} (\bibinfo {year} {2018})}\BibitemShut {NoStop}%
\bibitem [{\citenamefont {Pandey}\ \emph {et~al.}(2020)\citenamefont {Pandey},
  \citenamefont {Claeys}, \citenamefont {Campbell}, \citenamefont
  {Polkovnikov},\ and\ \citenamefont {Sels}}]{pandey_2020_adiabatic}%
  \BibitemOpen
  \bibfield  {author} {\bibinfo {author} {\bibfnamefont {M.}~\bibnamefont
  {Pandey}}, \bibinfo {author} {\bibfnamefont {P.~W.}\ \bibnamefont {Claeys}},
  \bibinfo {author} {\bibfnamefont {D.~K.}\ \bibnamefont {Campbell}}, \bibinfo
  {author} {\bibfnamefont {A.}~\bibnamefont {Polkovnikov}},\ and\ \bibinfo
  {author} {\bibfnamefont {D.}~\bibnamefont {Sels}},\ }\href
  {https://doi.org/10.1103/PhysRevX.10.041017} {\bibfield  {journal} {\bibinfo
  {journal} {Phys. Rev. X}\ }\textbf {\bibinfo {volume} {10}},\ \bibinfo
  {pages} {041017} (\bibinfo {year} {2020})}\BibitemShut {NoStop}%
\bibitem [{\citenamefont {Santos}\ \emph {et~al.}(2020)\citenamefont {Santos},
  \citenamefont {{P{\'e}rez-Bernal}},\ and\ \citenamefont
  {{Torres-Herrera}}}]{santos_2020_speck}%
  \BibitemOpen
  \bibfield  {author} {\bibinfo {author} {\bibfnamefont {L.~F.}\ \bibnamefont
  {Santos}}, \bibinfo {author} {\bibfnamefont {F.}~\bibnamefont
  {{P{\'e}rez-Bernal}}},\ and\ \bibinfo {author} {\bibfnamefont {E.~J.}\
  \bibnamefont {{Torres-Herrera}}},\ }\href
  {https://doi.org/10.1103/PhysRevResearch.2.043034} {\bibfield  {journal}
  {\bibinfo  {journal} {Phys. Rev. Research}\ }\textbf {\bibinfo {volume}
  {2}},\ \bibinfo {pages} {043034} (\bibinfo {year} {2020})}\BibitemShut
  {NoStop}%
\bibitem [{\citenamefont {Odavi\ifmmode~\acute{c}\else \'{c}\fi{}}\ \emph
  {et~al.}(2025)\citenamefont {Odavi\ifmmode~\acute{c}\else \'{c}\fi{}},
  \citenamefont {Viscardi},\ and\ \citenamefont
  {Hamma}}]{betheanzatsmagic2025}%
  \BibitemOpen
  \bibfield  {author} {\bibinfo {author} {\bibfnamefont {J.}~\bibnamefont
  {Odavi\ifmmode~\acute{c}\else \'{c}\fi{}}}, \bibinfo {author} {\bibfnamefont
  {M.}~\bibnamefont {Viscardi}},\ and\ \bibinfo {author} {\bibfnamefont
  {A.}~\bibnamefont {Hamma}},\ }\href {https://doi.org/10.1103/y9r6-dx7p}
  {\bibfield  {journal} {\bibinfo  {journal} {Phys. Rev. B}\ }\textbf {\bibinfo
  {volume} {112}},\ \bibinfo {pages} {104301} (\bibinfo {year}
  {2025})}\BibitemShut {NoStop}%
\bibitem [{\citenamefont {Collins}\ \emph {et~al.}(2022)\citenamefont
  {Collins}, \citenamefont {Matsumoto},\ and\ \citenamefont
  {Novak}}]{collins_2022_weingarten}%
  \BibitemOpen
  \bibfield  {author} {\bibinfo {author} {\bibfnamefont {B.}~\bibnamefont
  {Collins}}, \bibinfo {author} {\bibfnamefont {S.}~\bibnamefont {Matsumoto}},\
  and\ \bibinfo {author} {\bibfnamefont {J.}~\bibnamefont {Novak}},\ }\href
  {https://doi.org/10.1090/noti2474} {\bibfield  {journal} {\bibinfo  {journal}
  {Notices Amer. Math. Soc.}\ }\textbf {\bibinfo {volume} {69}},\ \bibinfo
  {pages} {1} (\bibinfo {year} {2022})}\BibitemShut {NoStop}%
\bibitem [{\citenamefont {Caux}\ and\ \citenamefont
  {Mossel}(2011)}]{caux_2011_remarks}%
  \BibitemOpen
  \bibfield  {author} {\bibinfo {author} {\bibfnamefont {J.-S.}\ \bibnamefont
  {Caux}}\ and\ \bibinfo {author} {\bibfnamefont {J.}~\bibnamefont {Mossel}},\
  }\href {https://doi.org/10.1088/1742-5468/2011/02/P02023} {\bibfield
  {journal} {\bibinfo  {journal} {J. Stat. Mech. Theor. Exp.}\ }\textbf
  {\bibinfo {volume} {2011}},\ \bibinfo {pages} {P02023} (\bibinfo {year}
  {2011})}\BibitemShut {NoStop}%
\end{thebibliography}

%

\end{document}